\documentclass[preprint]{aastex}
\usepackage[usenames,dvipsnames]{color}
\usepackage{natbib,graphicx,latexsym,lscape,pdflscape,url}
\usepackage[breaklinks,colorlinks,urlcolor=blue,citecolor=black,linkcolor=blue]{hyperref}

\newcommand{\Msun}{\mbox{$M_{\sun}$}}

\newcommand{\Lsun}{\mbox{$L_{\sun}$}}

\newcommand{\Mjup}{\mbox{$M_{\rm Jup}$}}
\newcommand{\Rjup}{\mbox{$R_{\rm Jup}$}}
\newcommand{\degree}{\mbox{$^{\circ}$}}
\newcommand{\perpix}{\mbox{pixel$^{-1}$}}

\newcommand{\masyr}{\hbox{mas\,yr$^{-1}$}}

\newcommand{\Kp}{\mbox{$K^{\prime}$}}
\newcommand{\Ks}{\mbox{$K_S$}}

\newcommand{\Mtot}{\mbox{$M_{\rm tot}$}}

\newcommand{\Lbol}{\mbox{$L_{\rm bol}$}}

\newcommand{\Teff}{\mbox{$T_{\rm eff}$}}
\newcommand{\logg}{\mbox{$\log{g}$}}

\newcommand{\Lp}{\mbox{${L^\prime}$}}

\newcommand{\obj}{SDSS~J1052+4422}
\newcommand{\objlong}{SDSS~J105213.51+442255.7}

\shorttitle{L/T Transition Mass--Luminosity Relation}
\shortauthors{Dupuy et al.}

\begin{document}

\title{The Mass--Luminosity Relation in the L/T Transition: \\
  Individual Dynamical Masses for the New $J$-Band Flux Reversal Binary \objlong{AB}\altaffilmark{*,\dag}}

\author{Trent J.\ Dupuy,\altaffilmark{1}
        Michael C.\ Liu,\altaffilmark{2}
        S.\ K.\ Leggett,\altaffilmark{3}
        Michael J.\ Ireland,\altaffilmark{4}
        Kuenley Chiu,\altaffilmark{5} \\ and
        David A.\ Golimowski\altaffilmark{6}}

      \altaffiltext{*}{Data presented herein were obtained at the
        W.M.\ Keck Observatory, which is operated as a scientific
        partnership among the California Institute of Technology, the
        University of California, and the National Aeronautics and
        Space Administration. The Observatory was made possible by the
        generous financial support of the W.M.\ Keck Foundation.}

      \altaffiltext{\dag}{Based on data obtained with WIRCam,
        a joint project of CFHT, Taiwan, Korea, Canada, France, at the
        Canada-France-Hawaii Telescope, which is operated by the
        National Research Council of Canada, the Institute
        National des Sciences de l'Univers of the Centre National de
        la Recherche Scientifique of France, and the University of
        Hawaii.}

      \altaffiltext{1}{The University of Texas at Austin, Department
        of Astronomy, 2515 Speedway C1400, Austin, TX 78712, USA}

      \altaffiltext{2}{Institute for Astronomy, University of Hawai`i,  
        2680 Woodlawn Drive, Honolulu, HI 96822, USA}

      \altaffiltext{3}{Gemini Observatory, Northern Operations Center,
        670 N.\ A'ohoku Place, Hilo, HI, 96720 USA}

      \altaffiltext{4}{Research School of Astronomy \& Astrophysics,
        Australian National University, Canberra ACT 2611, Australia}

      \altaffiltext{5}{C3 Energy, 1300 Seaport Boulevard Suite 500,
        Redwood City, CA 94063, USA}

      \altaffiltext{6}{Space Telescope Science Institute, 3700 San
        Martin Drive, Baltimore, MD 21218, USA}

\begin{abstract}

  We have discovered that \objlong\ (T$0.5\pm1.0$) is a binary in Keck
  laser guide star adaptive optics imaging, displaying a large
  $J$-to-$K$-band flux reversal ($\Delta{J} = -0.45\pm0.09$\,mag,
  $\Delta{K} = 0.52\pm0.05$\,mag).  We determine a total dynamical
  mass from Keck orbital monitoring ($88\pm5$\,\Mjup) and a mass ratio
  by measuring the photocenter orbit from CFHT/WIRCam absolute
  astrometry ($M_B/M_A = 0.78\pm0.07$).  Combining these provides the
  first individual dynamical masses for any field L or T dwarfs,
  $49\pm3$\,\Mjup\ for the L$6.5\pm1.5$ primary and $39\pm3$\,\Mjup\
  for the T$1.5\pm1.0$ secondary. Such a low mass ratio for a nearly
  equal luminosity binary implies a shallow mass--luminosity relation
  over the L/T transition ($\Delta\log\Lbol/\Delta\log{M} =
  0.6^{+0.6}_{-0.8}$).  This provides the first observational support
  that cloud dispersal plays a significant role in the luminosity
  evolution of substellar objects.  Fully cloudy models fail our
  coevality test for this binary, giving ages for the two components
  that disagree by 0.2\,dex (2.0$\sigma$).  In contrast, our observed
  masses and luminosities can be reproduced at a single age by
  "hybrid" evolutionary tracks where a smooth change from a cloudy to
  cloudless photosphere around 1300 K causes slowing of luminosity
  evolution.  Remarkably, such models also match our observed $JHK$
  flux ratios and colors well.  Overall, it seems that the
  distinguishing features \obj{AB}, like a $J$-band flux reversal and
  high-amplitude variability, are normal for a field L/T binary caught
  during the process of cloud dispersal, given that the age
  ($1.11^{+0.17}_{-0.20}$\,Gyr) and surface gravity ($\logg =
  5.0$--5.2) of \obj{AB} are typical for field ultracool dwarfs.

\end{abstract}


%----------------------------------------------------------------------%

\section{Introduction}

Perhaps the most important but uncertain process to model for
substellar objects is the formation, growth, and dispersal of
condensate clouds.  When present in the photosphere, clouds are a
dominant opacity source and thereby control basic observable
properties like broadband colors, magnitudes, and spectra.  In brown
dwarfs, clouds appear to grow in influence going from early to
late-type L dwarfs then begin dispersing with early-type T dwarfs,
resulting in drastic changes in near-infrared spectra.  The most
prominent features of this transition are that $JHK$ colors become
significantly bluer \citep[e.g.,][]{2002ApJ...564..452L} and $J$-band
fluxes become brighter \citep[e.g.,][]{2002AJ....124.1170D} going from
late L to early T dwarfs.  Early analysis of such observations
indicated that the L/T transition occupies a narrow range of effective
temperature (\Teff), although the underlying physical process
explaining the brightening at $J$ band was debated.  For example,
rapid changes in cloud sedimentation efficiency
\citep{2004AJ....127.3553K} or breakup caused by convection
\citep{2002ApJ...564..421B} could reproduce the $J$-band brightening
at a single \Teff, while \citet{2003ApJ...585L.151T} suggested that
mass/age spreads in the population of field brown dwarfs were
responsible, not changes in the cloud themselves.

The discovery of binaries at the L/T transition in which one component
was directly observed to be brighter at 1.0--1.3\,\micron\ ($Y$ or
$J$) but fainter at other wavelengths \citep[$I$, $H$, or
$K$;][]{2003AJ....125.3302G, 2006ApJ...647.1393L} provided the first
unambiguous evidence that the $J$-band brightening must occur along a
single isochrone.  Such flux reversals require a significant flux
redistribution as brown dwarfs cool, most likely brought on by changes
in cloud opacity.  Recent discoveries of large-amplitude variables at
near-infrared wavelengths, so far only reliably detected in the L/T
transition \citep[e.g.,][]{2009ApJ...701.1534A, 2012ApJ...750..105R,
  2014ApJ...793...75R}, support the idea that cloud clearing is
spatially heterogeneous in the photosphere. However, without well
determined masses and/or ages for any systems that display a $J$-band
flux reversal or weather, the alternative possibility of unusual cloud
properties, e.g., due to surface gravity, exists. In fact, there still
remains only one object in the L/T transition with a precise dynamical
mass measurement \citep[LHS~2397aB, photometric spectral type estimate
of L7;][]{2009ApJ...699..168D}. 

For nearly all substellar objects found to date, evolutionary models
are the sole means for estimating their physical properties, typically
by using the observed luminosity and an adopted age to yield a
model-dependent mass (and temperature, radius, and surface
gravity). Such models require an assumption for the photosphere
opacity as a boundary condition, and one of the key challenges is the
treatment of condensate clouds.  The formation, growth, and settling
of dust condensates likely occurs at many different levels in the
atmosphere and would thus also be influenced both by the local
physical conditions and the bulk motions of the gas via convection
\citep{2010A&A...513A..19F}. There are numerous approaches to modeling
these complex processes and parameterizing them so that they can be
incorporated into one-dimensional atmospheric models \citep[e.g., see
recent reviews by][]{2014A&ARv..22...80H, 2014arXiv1410.6512M}.
However, the currently available evolutionary models assume one of two
limiting cases for treatment of the dust.  Either dust exists in
chemical equilibrium with the gas, resulting in thicker clouds at
cooler temperatures \citep{2000ApJ...542..464C}, or the grains rapidly
fall out of the photosphere as soon as they form, leaving behind
dust-free gas \citep{1997ApJ...491..856B, 2003A&A...402..701B}.
Attempting to match the observations described above that dust clouds
disperse over a narrow range of \Teff, \citet{2008ApJ...689.1327S}
computed evolutionary models where the atmosphere is interpolated
between the fully cloudy and cloud free limiting cases as objects cool
from 1400\,K to 1200\,K.  Despite the limitations of these various
simplifying assumptions, current models are at least broadly in accord
with the observed substellar sequence in open clusters
\citep[e.g.,][]{2014MNRAS.445.3908L, 2015arXiv150203728B} and in the
field \citep[e.g.,][]{2003AJ....126..975T, 2008ApJ...689.1327S},
although discrepancies are obvious in regimes where photospheric
condensates play a more significant role, especially the L/T
transition.

More stringent test of the theoretical models are now within
reach. The past decade saw a growing number of substellar visual
binaries with dynamical masses measured via astrometric monitoring
\citep[e.g.,][]{2001ApJ...560..390L, 2004A&A...423..341B,
  2008ApJ...689..436L, 2009ApJ...706..328D, 2009ApJ...692..729D,
  2009ApJ...699..168D, 2010ApJ...721.1725D, 2010ApJ...711.1087K}.  The
most powerful tests to date come from brown dwarf binaries in a
hierarchical triple with a main-sequence star, where the subtellar
binary orbit gives its dynamical total mass and the primary star gives
the system age from gyrochronology. For the two known systems where
this is possible, the models seem to predict luminosities that are
systematically 0.2--0.4\,dex lower than observed
\citep{2009ApJ...692..729D, 2014ApJ...790..133D}. However, without
individual masses the mass--luminosity relation is unconstrained, and
thus a complementary test would be to obtain masses and luminosities
for a coeval binary system, even in the absence of an age
determination.  Previous work has resulted in individual masses for
late-M dwarfs, showing broad agreement with the mass--luminosity
relation as models approach the substellar
boundary.\footnote{Individual masses have been determined for two
  field late-M dwarf systems: one from resolved radial velocities and
  relative astrometry
  \citep[Gl~569Bab;][]{2004astro.ph..7334O,2006ApJ...644.1183S,2010ApJ...711.1087K},
  and one from absolute astrometry
  \citep[LHS~1070BC;][]{2008A&A...484..429S,2012A&A...541A..29K}.  The
  young brown dwarf eclipsing binary 2MASS~J05352184$-$0546085AB
  (M6+M6) in the Orion Nebula Cluster also has well determined
  individual masses \citep{2006Natur.440..311S}.}  Further tests of
evolutionary models are sorely needed, especially in the L/T
transition where they are routinely employed to characterize
planetary-mass discoveries, e.g., 2MASSW~J1207334$-$393254b
\citep{2004A&A...425L..29C}, HR~8799bcde \citep{2008Sci...322.1348M,
  2010Natur.468.1080M}, and PSO~J318.5338$-$22.8603
\citep{2013ApJ...777L..20L}.

We present the discovery of a new $J$-band flux reversal binary,
\objlong{AB} (hereinafter \obj{AB}), along with high-precision
dynamical masses of the individual components based on resolved
orbital monitoring from Keck and absolute astrometry from the
Canada-France-Hawaii Telescope (CFHT). \obj\ was originally discovered
by \citet{2006AJ....131.2722C}. They determined an integrated-light
spectral type of T$0.5\pm1.0$ according to the
\citet{2006ApJ...637.1067B} indices (T$0.0\pm1.0$ according to
\citealp{2002ApJ...564..466G} indices). More recently,
\citet{2013ApJ...767...61G} found that \obj\ is a high-amplitude
variable with peak-to-peak variations of up to 0.06\,mag in $J$ band,
although their integrated-light observations could not determine which
component was responsible for the variability. Our mass determination
for \obj{AB} is therefore the first for a $J$-band flux reversal
binary and the first for a brown dwarf displaying significant weather.
More generally, our results are also the first individual mass
measurements for any field L or T dwarfs.  This is distinct from the
aforementioned results on dynamical total masses, as the only
individual masses in this spectral type range are for two substellar
companions to stars measured from
absolute astrometry \citep[Gl~802B;][]{2008ApJ...678..463I},
or relative astrometry combined with radial velocities
\citep[HR~7672B;][]{2012ApJ...751...97C}.
There are also a number of stellar model-dependent mass determinations
for brown dwarfs in eclipsing systems \citep{2008A&A...491..889D,
  2011ApJ...726L..19A, 2011A&A...533A..83B, 2011A&A...525A..68B,
  2011ApJ...730...79J, 2012ApJ...761..123S, 2013A&A...551L...9D,
  2014arXiv1411.4047M}. However, all of these companions lack the
spectral information available for field L and T dwarfs that enables
the strongest tests of substellar models.

%----------------------------------------------------------------------%

\section{Discovery and Astrometric Monitoring of \obj{AB} \label{sec:obs}}

\subsection{Keck/NIRC2 LGS AO \label{sec:keck}}

We observed \obj\ on 2005~May~1~UT with the then recently commissioned
laser guide star adaptive optics (LGS AO) system at the Keck~II
telescope \citep{2004SPIE.5490..321B, 2006PASP..118..297W,
  2006PASP..118..310V}.  We used the facility near-infrared camera
NIRC2, obtaining five dithered images in \Kp\ band.  \obj\ appeared to
be marginally resolved (peanut shaped) in these images, indicating
that it was likely a binary.  In follow-up imaging on 2006~May~5~UT,
\obj\ was more obviously resolved because it had moved to a wider
projected separation of 70\,mas, as compared to 42\,mas in 2005.  We
obtained data in the Mauna Kea Observatories (MKO) $J$, $H$, and \Ks\
photometric bandpasses \citep{2002PASP..114..169S,
  2002PASP..114..180T} and discovered that while the western component
was brighter in \Ks\ and perhaps $H$ band, the eastern component was
in fact brighter in $J$ band.  In keeping with the convention with
previous $J$-band flux reversal binaries
\citep[e.g.,][]{2006ApJ...647.1393L, 2008ApJ...685.1183L}, we will
refer to the component brighter in \Ks-band as the primary (\obj{A}).

We continued Keck AO monitoring of \obj{AB} in order to determine its
orbital parameters and thereby a total dynamical mass.  Our
observations are a combination of normal imaging and data taken with
the 9-hole non-redundant aperture mask installed in the filter wheel
of NIRC2 \citep{2006SPIE.6272E.103T}.
On some nights we obtained data using the natural guide star (NGS) AO
system, because the tip-tilt star is bright enough ($R \approx
14.6$\,mag) and close enough to the target (19$\arcsec$ away) that it
can sometimes be used as an NGS.  The analysis of our data was the
same regardless of whether we observed in NGS or LGS mode.

Our procedure for reducing and analyzing NIRC2 imaging data is
described in detail in our previous work \citep{2009ApJ...706..328D,
  2009ApJ...692..729D, 2009ApJ...699..168D, 2010ApJ...721.1725D}. To
summarize briefly, we measure binary parameters using a
three-component Gaussian representation of the point-spread function.
We derive uncertainties by applying our fitting method to artificial
binary images constructed from images of single stars with similar
full-width half-maxima (FWHM) and Strehl ratios, as well as by
checking the scatter between individual dithered images.  We use the
NIRC2 astrometric calibration from \citet{2010ApJ...725..331Y}, which
includes a correction for the nonlinear distortion of the camera and
has a pixel scale of $9.952\pm0.002$\,mas\,\perpix\ and an orientation
for the detector's $+y$-axis of $+0\fdg252\pm0\fdg009$ east of north.
Analysis of our masking data was performed using a pipeline similar to
that used in previous papers containing NIRC2 masking data
\citep[e.g.,][]{2008ApJ...678..463I, 2008ApJ...678L..59I} and is
described in detail in Section~2.2 of \citet{2009ApJ...699..168D}.

\begin{figure}

\centerline{
\includegraphics[height=1.2in,angle=0]{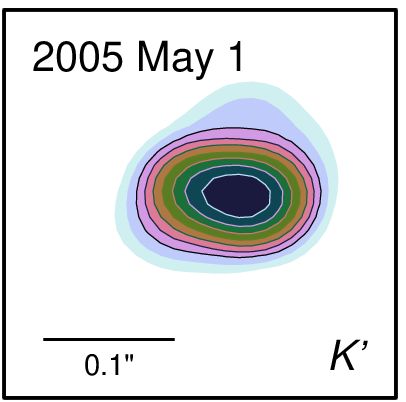} \hskip 0.04in
\includegraphics[height=1.2in,angle=0]{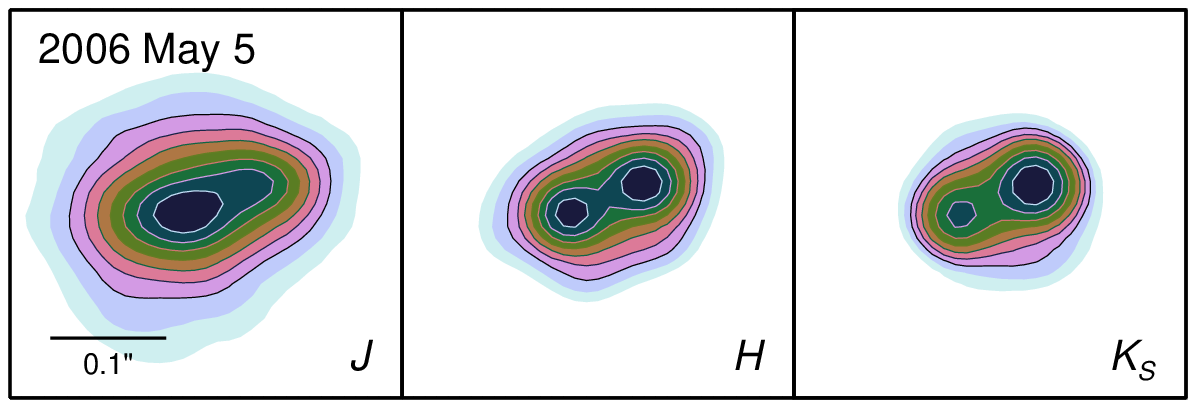} \hskip 0.04in
\includegraphics[height=1.2in,angle=0]{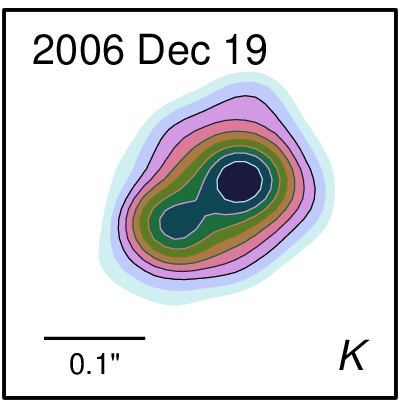} } 
\vskip 0.05in 
\centerline{
\includegraphics[height=1.2in,angle=0]{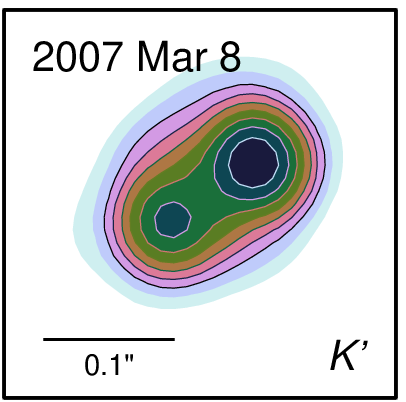} \hskip 0.04in
\includegraphics[height=1.2in,angle=0]{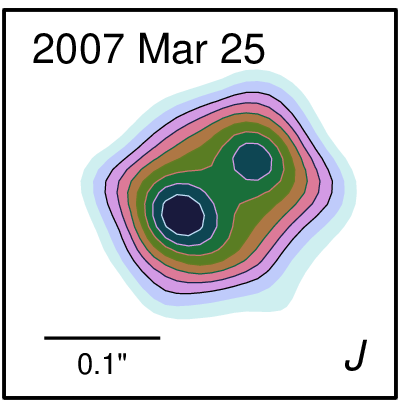} \hskip 0.04in
\includegraphics[height=1.2in,angle=0]{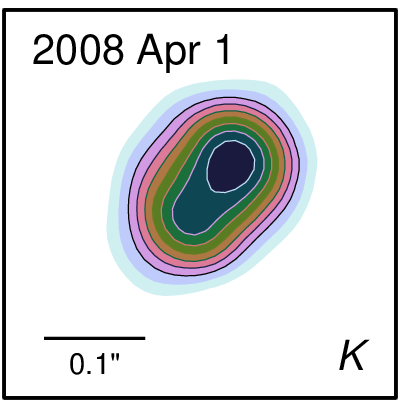} \hskip 0.04in
\includegraphics[height=1.2in,angle=0]{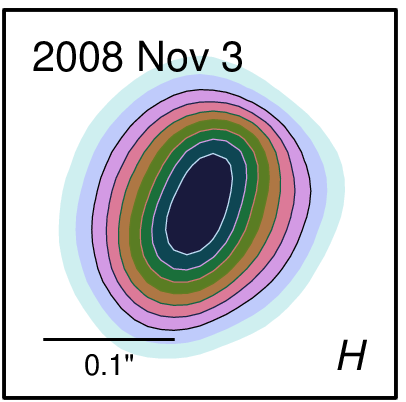} }
\vskip 0.3in

\caption{\normalsize Contour plots of our Keck AO images from which we
  derive astrometry and flux ratios (Table~\ref{tbl:keck}).  Contours
  are in logarithmic intervals from unity to 10\% of the peak flux in
  each band.  The image cutouts are all the same size and have the
  same native pixel scale, and we have rotated them such that north is
  up for display purposes. \label{fig:keck}}

\end{figure}

\begin{figure}

\centerline{
\includegraphics[height=2.0in,angle=0]{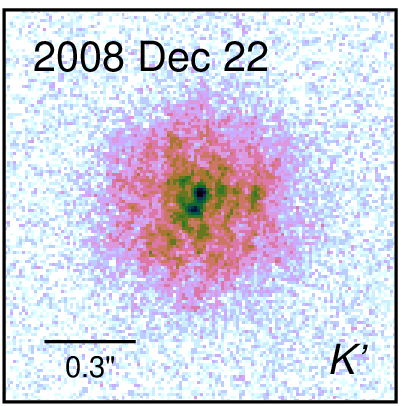} \hskip 0.04in
\includegraphics[height=2.0in,angle=0]{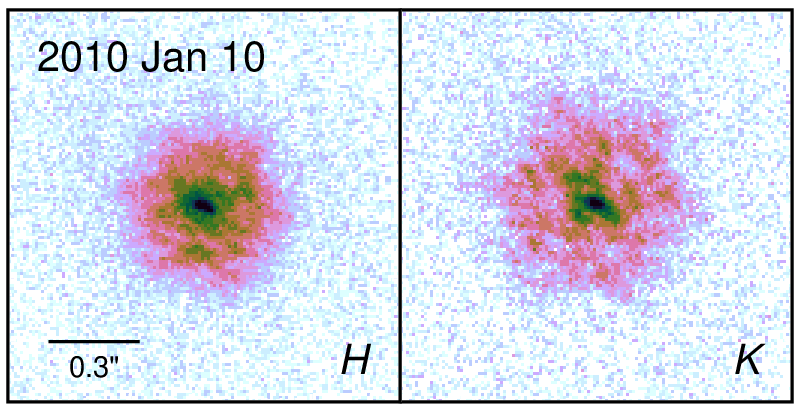} } 
\vskip 0.05in 
\centerline{
\includegraphics[height=2.0in,angle=0]{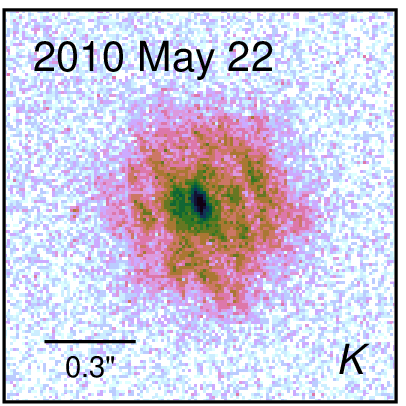} \hskip 0.04in
\includegraphics[height=2.0in,angle=0]{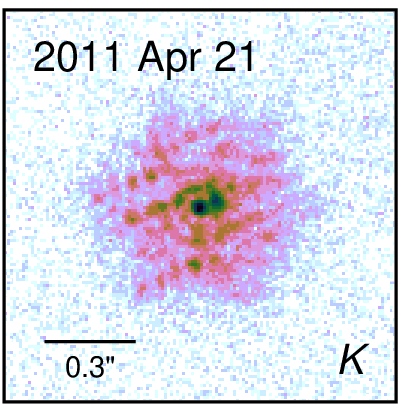} \hskip 0.04in
\includegraphics[height=2.0in,angle=0]{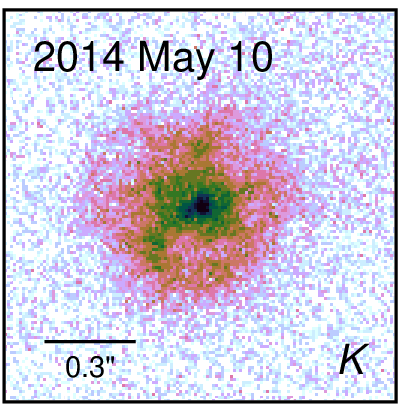} }
\vskip 0.3in

\caption{\normalsize Keck/NIRC2 images of the interferograms produced
  when observing \obj{AB} with the 9-hole aperture mask.  The binary
  can be seen by eye as an elongation or double peak in the center of
  the point-spread function in all but one epoch.  In data from
  2010~May~22~UT the binary is very tight ($39.9\pm0.7$\,mas), and the
  elongation is instead along the elevation axis (205\degree) caused
  by atmospheric dispersion given the modest airmass (1.32) of the
  observation.  These image cutouts are all the same size, have the
  same native pixel scale, have been rotated such that north is up,
  and are shown with a square-root stretch.  \label{fig:mask}}

\end{figure}

A summary of our Keck AO observations is given in
Table~\ref{tbl:keck}, including the binary separation, position angle
(PA), and flux ratio as well as the FWHM and Strehl ratio of our AO
images at each epoch.  Contour plots of our imaging data are shown in
Figure~\ref{fig:keck}, and images of our masking interferograms are
shown in Figure~\ref{fig:mask}.  At the two epochs where we have data
in more than one bandpass our derived binary parameters are consistent
within the errors.  However, unlike previous binaries we have
monitored \citep[e.g., Gl~417BC;][]{2014ApJ...790..133D}, our flux
ratios for \obj{AB} are not always consistent between epochs.  The
small variations we observe at $J$ and $H$ bands are consistent with
variability at the $\approx$0.10\,mag level as implied by the
integrated-light variability of 0.06\,mag reported by
\citet{2013ApJ...767...61G}.  In the following analysis we use the
weighted average flux ratio for each bandpass, assuming an additional
0.10\,mag error added in quadrature to account for variability.  This
gives $\Delta{J} = -0.45\pm0.09$\,mag, $\Delta{H} = 0.06\pm0.07$\,mag,
and $\Delta{K} = 0.52\pm0.05$\,mag.  The only other ultracool binary
known to have such a large $J$-band flux reversal is
2MASS~J14044948$-$3159330AB \citep[$\Delta{J} =
-0.54\pm0.08$\,mag;][]{2008ApJ...685.1183L, 2012ApJS..201...19D}.

\subsection{CFHT/WIRCam Astrometry \label{sec:cfht}}

We have been monitoring \obj{AB} as part of the Hawaii Infrared
Parallax Program at the CFHT in order to measure the precise distance
needed for a dynamical mass determination. Our methods for obtaining
high-precision astrometry from the facility near-infrared wide-field
imager WIRCam \citep{2004SPIE.5492..978P} are described in detail in
\citet{2012ApJS..201...19D}. We have obtained a total of 427 $J$-band
images centered on \obj{AB} over 21 epochs spanning 6.79\,yr. At each
epoch, we measured the position of \obj{AB} in integrated light along
with 30 other stars in the field having signal-to-noise ratios (S/N)
$>$23. The subset of 26 stars that appear in the SDSS-DR9 catalog
\citep{2012ApJS..203...21A} were used for the absolute astrometric
calibration of the linear terms (pixel scales in $x$ and $y$,
rotation, and shear).  We simultaneously fit for the proper motion and
parallax of all stars in the field and found no other sources
co-moving with \obj{AB} down to $J=18.2$\,mag. The median and rms of
the seeing was $0\farcs59\pm0\farcs09$ over our observations, and the
S/N of \obj{AB} ranged from 80--160.  The absolute positions of
\obj{AB} measured from our CFHT data are listed in
Table~\ref{tbl:cfht}.

Given the 6.8\,yr time baseline of our CFHT observations, and the
8.6\,yr orbital period of \obj{AB} that we determine from our Keck
astrometry (Section~\ref{sec:orbit}), significant orbital motion might
be expected to be observed in our integrated-light astrometry. Indeed,
we saw an orbital arc in our CFHT residuals that caused a very poor
$\chi^2 = 428.8$ (37 degrees of freedom) in our initial parallax and
proper motion fit to the data. Thus, we must combine our resolved
orbital analysis from Keck with our CFHT astrometry in order to
accurately retrieve the parallax of \obj{AB}.

%----------------------------------------------------------------------%

\section{Measured Properties of \obj{AB} \label{sec:prop}}

\subsection{Orbital Parameters \& Parallax \label{sec:orbit}}

We performed a joint analysis of our two astrometric data sets for
\obj{AB}: resolved measurements from Keck AO and integrated-light
positions from CFHT/WIRCam.  All but one of the seven visual binary
orbit parameters are shared in common between the Keck and CFHT data.
Since our Keck data only gives us the position of one binary component
relative to the other, we fit for the total semimajor axis ($a = a_1 +
a_2$) that is the sum of the individual component's semimajor axes
about the center of mass.  In our CFHT data, we only see the motion of
the photocenter, the amplitude of which depends on the flux ratio and
mass ratio of the binary.  We therefore fit for a photocenter
semimajor axis ($\alpha$) that we will later use to derive the system
mass ratio.  We also fit for the five usual parameters needed for our
CFHT parallax data: R.A.\ zero point and proper motion, Dec.\ zero
point and proper motion, and parallax.  Therefore, there are a total
of 13 parameters in the joint fit of our Keck and CFHT data.

To determine probability distributions for the orbit and parallax
parameters, we performed a Markov Chain Monte Carlo (MCMC) analysis.
Unlike our previous work, we used the Python implementation of the
affine-invariant ensemble sampler \texttt{emcee~v2.1.0}
\citep{2013PASP..125..306F}.  This allows for more efficient
exploration of our 13-dimensional parameter space than our own custom
MCMC tools that used a Metropolis-Hastings jump acceptance criterion
with Gibbs sampling.  We adopted uniform priors in the logarithms of
period and semimajor axis ($\log{P}$, $\log{a}$), eccentricity ($e$),
argument of periastron ($\omega$), PA of the ascending node
($\Omega$), mean longitude at a reference time ($\lambda_{\rm ref}$),
and the ratio of the photocenter semimajor axis to the total semimajor
axis ($\alpha/a$).  The reference time is $t_{\rm ref} = 2455197.5$~JD
(2010~Jan~1 00:00~UT), which is related to the time of periastron
passage $T_0=t_{\rm ref} - P\times(\lambda_{\rm ref} - \omega) /
360$\degree.  We assume randomly distributed viewing angles by
adopting an inclination prior uniform in $\cos{i}$.  We adopt uniform
priors in the proper motion and R.A.\ and Dec.\ zero points and a
uniform spatial volume prior in the parallax.  The latter is
justifiable as \obj\ was discovered well above the magnitude limits of
the SDSS survey ($m_{\rm lim} - M \approx 3.5$\,mag).  The effect of
this discovery bias on the parallax prior was considered by
\citet[][see their Figure~S3]{2013Sci...341.1492D}, and we find that
\obj\ would be well within the uniform space density regime.
Regardless, we note that because of the high precision of our measured
parallax this choice of prior has an indistinguishable effect on the
credible interval of this parameter ($\leq$0.1\%).  In other words,
the observational constraints dominate over the prior in determining
the posterior distribution of the parallax. We used 10$^3$ walkers of
10$^4$ steps each, saving only every hundredth step and discarding the
first 10\% of steps as the burn-in time for each walker.

\begin{figure}
\centerline{
\includegraphics[width=4.0in,angle=0]{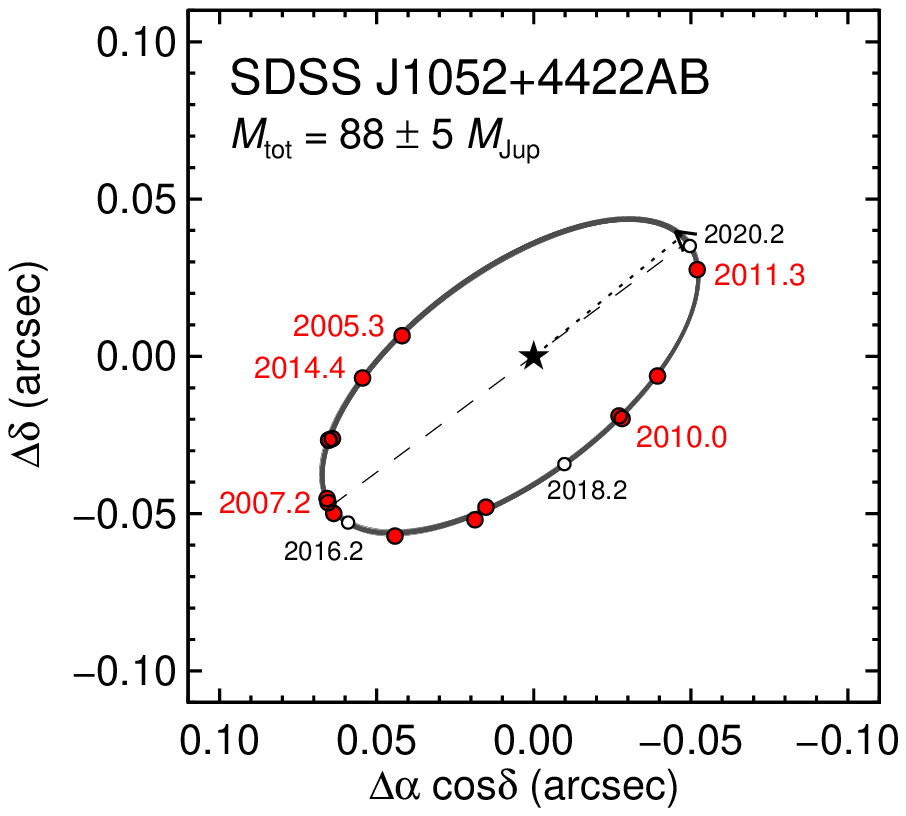}\hskip  0.2in
\includegraphics[width=2.3in,angle=0]{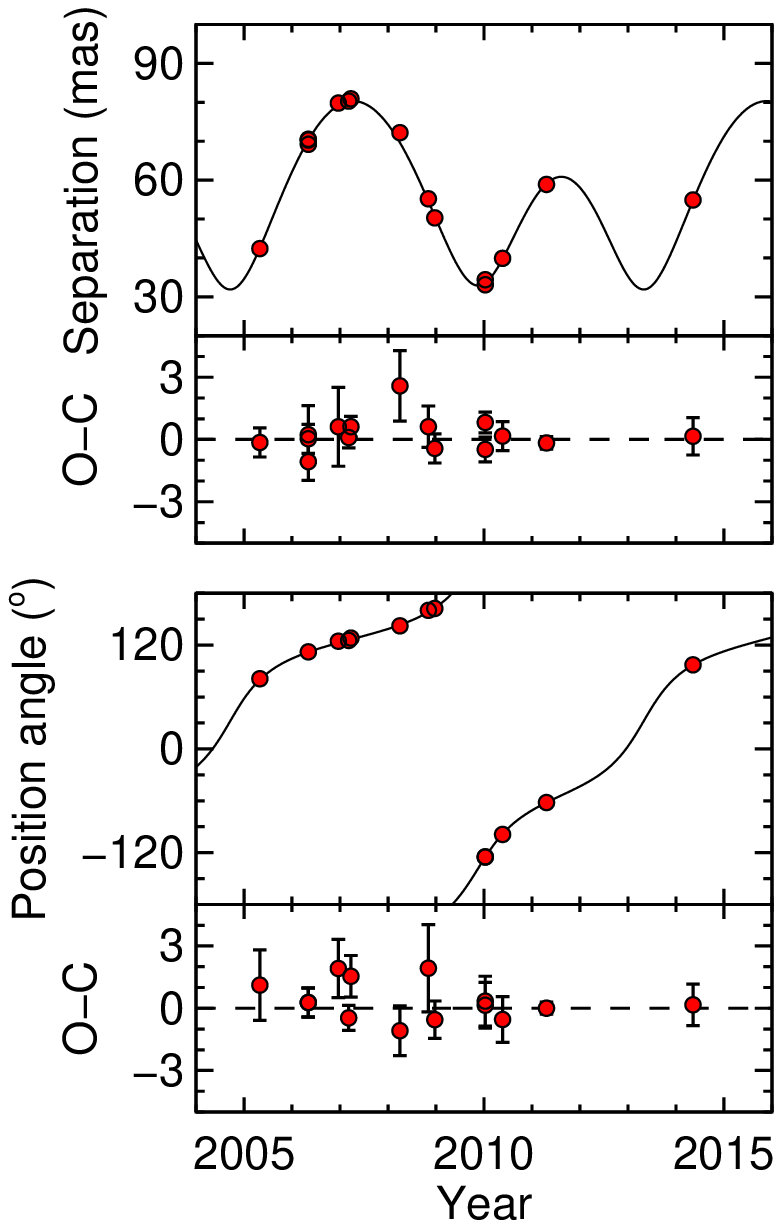}}

\caption{\normalsize \emph{Left:} Keck AO relative astrometry for \obj{AB}
  along with 100 randomly drawn orbits from our MCMC analysis
  individually plotted as thin lines. Error bars for the data points
  are smaller than the plotting symbols. The short dotted line
  indicates the time of periastron passage, the long dashed line shows
  the line of nodes, and small empty circles show predicted future
  locations.  \emph{Right:} Measurements of the projected separation
  and PA of \obj{AB}.  The best-fit orbit is shown as a solid line.
  The bottom panels show the observed minus computed ($O-C$)
  measurements with observational error bars.\label{fig:orbit}}

\end{figure}

The best-fit parameters and credible intervals derived from our MCMC
posterior distributions are given in Table~\ref{tbl:orbit}.  We found
an orbital period of $8.608\pm0.025$\,yr (0.29\% error) and total
semimajor axis of $70.67\pm0.24$\,mas (0.34\% error), and accounting
for the slight covariance between these parameters results in an
uncertainty in the dynamical total mass of 1.1\% from our orbit
determination alone.  As a check on our new MCMC methods, we performed
a separate MCMC analysis on just the Keck data using our own
Metropolis-Hastings code \citep{2014ApJ...790..133D}.  The resulting
1$\sigma$ credible intervals for the seven visual binary parameters
were consistent to within a fraction of 1$\sigma$.  The resolved orbit
of \obj{AB} is shown in Figure~\ref{fig:orbit} along with our Keck
astrometry.

\begin{figure}
\centerline{
\hskip -0.2in
\includegraphics[width=2.1in,angle=0]{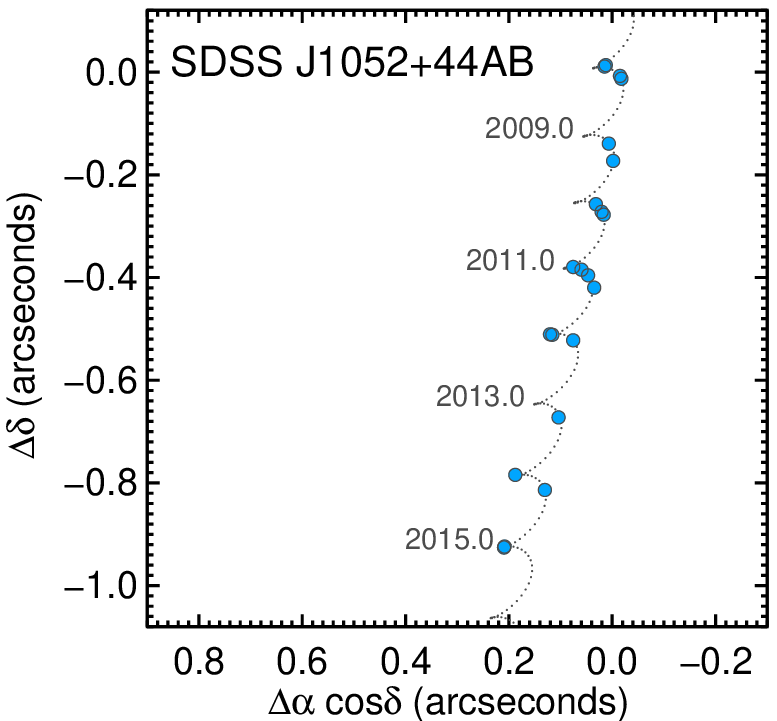} \hskip  0.05in
\includegraphics[width=2.1in,angle=0]{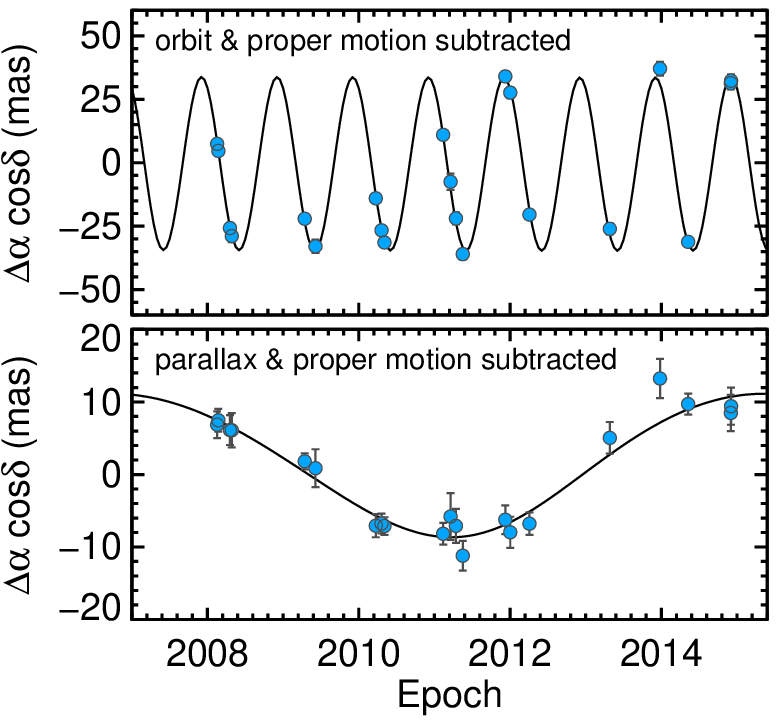} \hskip  0.05in
\includegraphics[width=2.1in,angle=0]{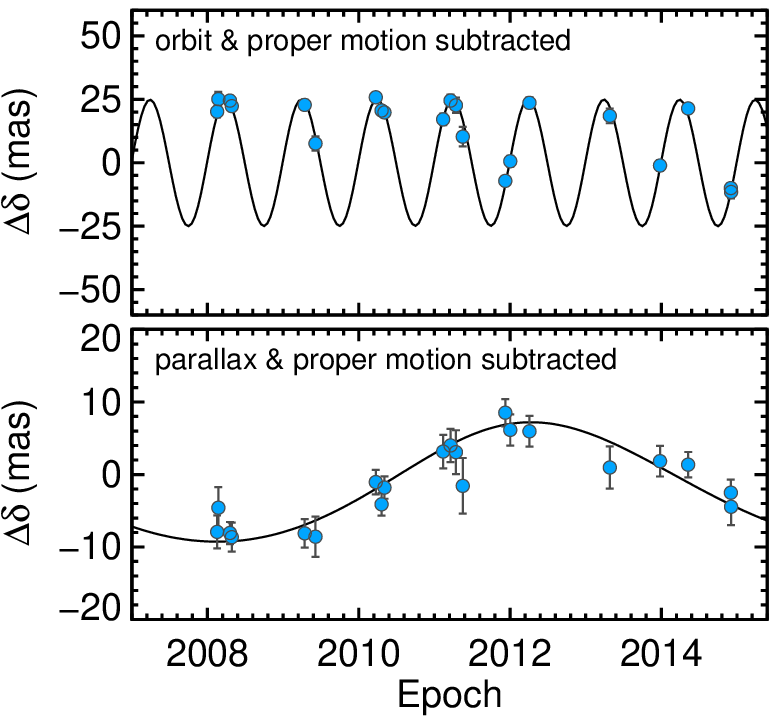}}

\caption{\normalsize \emph{Left:} CFHT/WIRCam integrated-light
  astrometry for \obj{AB} (blue circles) along with the best-fit model
  incorporating proper motion, parallax, and photocenter orbital
  motion (dotted line).  \emph{Middle, Right:} The same astrometry
  except with the best-fit proper motion and orbital motion removed,
  leaving just the parallax (top), and with the best-fit proper motion
  and parallax removed, leaving just the orbital motion of the
  photocenter (bottom). Error bars are plotted on all panels, but they
  are typically only visible in the plots displaying orbital
  motion. \label{fig:plx}}

\end{figure}

The additional parameters we fitted to our integrated-light CFHT data
provide the proper motion and parallax relative to our grid of
astrometric reference stars, as well as the size of the photocenter's
orbit\footnote{We quote the photocenter semimajor axis as a negative
  value because the photocenter motion is the opposite of what is seen
  in typical pairings of stars, brown dwarfs, or planets.  Normally,
  the less massive component is fainter and thus the center-of-light
  follows the brighter, more massive component's motion.  In the case
  of \obj{AB}, the center of $J$-band light follows the secondary
  component.  This can be seen when comparing Figures~\ref{fig:orbit}
  and \ref{fig:plx} where, e.g., in 2007 the secondary is seen in Keck
  data to be southeast of the primary and in CFHT data the photocenter
  shift is also to the southeast.} ($\alpha = -11.6\pm0.6$\,mas). This
best-fit solution is shown in Figure~\ref{fig:plx}. In order to
compute the distance, we derived a correction to account for the mean
parallax of our reference grid from the Besan\c{c}on model of the
Galaxy \citep{2003A&A...409..523R}. We found $\pi_{\rm abs} - \pi_{\rm
  rel} = 1.7\pm0.3$\,mas, where the uncertainty corresponds to the
statistical variance in sampling 30 stars in the $J$-band magnitude
range of our images, according to the much larger modeled Besan\c{c}on
population.  Adding this to the relative parallax results in an
absolute parallax of $38.4\pm0.7$\,mas, corresponding to a distance of
$26.1\pm0.5$\,pc.  Similarly, we computed additive corrections to our
proper motions of $\Delta{\mu_{\rm R.A.}} = -6\pm3$\,\masyr\ and
$\Delta{\mu_{\rm Dec.}}  = -7\pm3$\,\masyr. As a check, we input our
absolute proper motion and parallax to the BANYAN~II v1.3 web tool
\citep{2013ApJ...762...88M, 2014ApJ...783..121G} but found no linkage
to the seven kinematic associations in their solar neighborhood model.

\subsection{Dynamical Masses \label{sec:mass}}

Combining our measured parallactic distance with the total semimajor
axis and orbital period gives a precise total system mass for \obj{AB}
of $88\pm5$\,\Mjup\ (6\% error). We can also compute the mass ratio
and thereby individual component masses by considering the photocenter
motion seen in our integrated-light CFHT data. We found the ratio of
the photocenter semimajor axis to the total semimajor axis was
$\alpha/a = -0.164\pm0.008$. This ratio is set by the flux ratio and
mass ratio of the binary, such that $\alpha/a = f - \beta$. The first
parameter is the ratio of the secondary's mass to the total mass, $f =
M_B/(M_A+M_B)$, and the second parameter is the ratio of secondary's
flux to the total flux, $\beta = L_B/(L_A+L_B)$. Our $J$-band flux
ratio measured from Keck is $\Delta{J} = -0.45\pm0.09$\,mag, which
corresponds to $\beta = 0.602\pm0.020$. Solving for $f$ gives
$0.438\pm0.022$ and thus a mass ratio of $q \equiv M_B/M_A =
0.78\pm0.07$. This in turn gives individual masses of $49\pm3$\,\Mjup\
for \obj{A} and $39\pm3$\,\Mjup\ for \obj{B}. Therefore, we validate
for the first time that assumed primary component in a $J$-band flip
system is indeed more massive, and the mass ratio is surprisingly low.
We also directly determine that both components are unambiguously
substellar \citep[$<$75\,\Mjup;][]{1997A&A...327.1039C}.

\subsection{Spectral Types \label{sec:spt}}

In order to fully characterize the \obj{AB} system and aid in
computing bolometric corrections for the components, we have
determined the component spectral types through decomposition of its
integrated-light spectrum. \citet{2008ApJ...681..579B} published a
SpeX prism spectrum of \obj\ in integrated light ($R = 120$) which we
obtained from the SpeX Prism
Libraries.\footnote{\url{http://pono.ucsd.edu/~adam/browndwarfs/spexprism}}
We performed spectral decomposition analysis using the method
described in Section~5.2 of \citet{2012ApJS..201...19D}. Briefly, we
started with all possible pairs of the 178 IRTF/SpeX prism spectra
from the library of \citet{2010ApJ...710.1142B}. For each template
pairing we determined the scale factors needed to minimize the
$\chi^2$ compared to our observed spectrum. This resulted in a set of
$J$-, $H$-, and $K$-band flux ratios for each pairing, which we
compared to the flux ratios we measured from our Keck AO images
($\Delta{J} = -0.45\pm0.09$\,mag, $\Delta{H} = 0.06\pm0.07$\,mag, and
$\Delta{K} = 0.52\pm0.05$\,mag). We excluded template pairs that
disagreed significantly with our measured flux ratios, $p(\chi_{\rm
  phot}^2) < 0.05$, and then examined the ensemble of template pairs
that provided the best spectral matches.

\begin{figure}

\centerline{\includegraphics[width=6.0in,angle=0]{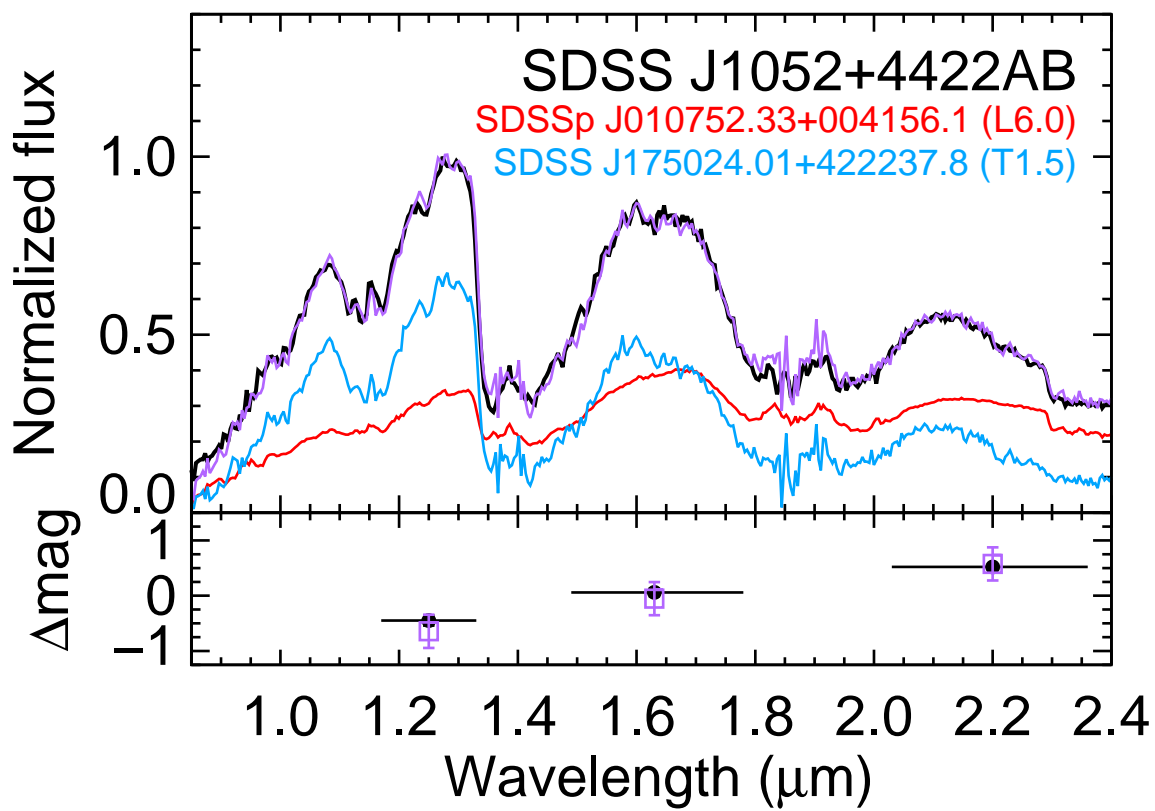}}

\caption{\normalsize Integrated-light spectrum of \obj{AB} (black) and
  best matching component templates (colored lines).  The bottom
  subpanel shows the observed $J$-, $H$-, and $K$-band broadband flux
  ratios used to constrain the decomposition (filled black circles
  with errors) and the resulting flux ratios computed from the best
  matching template pair (open colored squares). \label{fig:spec}}

\end{figure}

The best match to our spectrum was provided by the templates
SDSSp~J010752.33+004156.1 (L6) and SDSS~J175024.01+422237.8 (T1.5),
where we use the infrared types reported by
\citet{2010ApJ...710.1142B}.  This best-fit spectral template match is
shown in Figure~\ref{fig:spec}.  The next best matches use primary
templates with types ranging from L4.5:: (2MASSW~J0820299+450031,
typed in the optical as L5 by \citealp{2000AJ....120..447K}) to L8.5
and secondary templates with types ranging from T0:
(SDSS~J015141.69+124429.6, typed in the infrared as T1 by
\citealp{2006ApJ...637.1067B}) to T2.5.  We therefore adopt types of
L$6.5\pm1.5$ for \obj{A} and T$1.5\pm1.0$ for \obj{B}.

\subsection{Bolometric Luminosities \label{sec:lbol}}

By combining our Keck flux ratios with published MKO system photometry
for \obj{AB} \citep{2006AJ....131.2722C} and our CFHT parallax, we are
able to estimate the component luminosities.  Given the fact that the
flux ratio flips between $J$ and $K$ bands, we first consider the
bolometric luminosity (\Lbol) implied by each bandpass separately. We
used the polynomial relations between spectral type and bolometric
correction (BC) from \citet{2010ApJ...722..311L}. To determine the
uncertainty in the bolometric correction we allow for spectral type
uncertainties in a Monte Carlo fashion, compute the rms, and then add
the published rms scatter about the polynomial relation in quadrature.
In $J$ band we find bolometric corrections of $1.50\pm0.16$\,mag and
$1.94\pm0.24$\,mag for the primary and secondary, respectively. This
BC difference exactly compensates for the fact that the secondary is
brighter in $J$ band, resulting in nearly identical luminosities of
$\log(\Lbol/\Lsun) = -4.62\pm0.07$\,dex and $-4.62\pm0.10$\,dex,
respectively. Similarly, in $H$ band where our photometry is
consistent with the two components having equal flux, the BC
compensates and gives $\log(\Lbol/\Lsun) = -4.59\pm0.04$\,dex and
$-4.64\pm0.04$\,dex. We find comparable results using $K$ band of
$\log(\Lbol/\Lsun) = -4.57\pm0.05$\,dex and $-4.63\pm0.06$\,dex.

We chose to use the luminosities derived from our $K$ band photometry
because it is the least likely to be affected by the variability
observed by \citet{2013ApJ...767...61G} in $J$ band, and we have many
more $K$-band flux ratio measurements than at $J$ or $H$ bands.  Our
$K$-band flux ratio has the smallest uncertainty, and the scatter in
the BC$_K$ relation (0.08\,mag) is almost as small as for BC$_H$
(0.07\,mag).  We note however that the \Lbol\ estimates in all bands
are consistent within the uncertainties.

\begin{figure}

\centerline{\includegraphics[height=3.5in,angle=0]{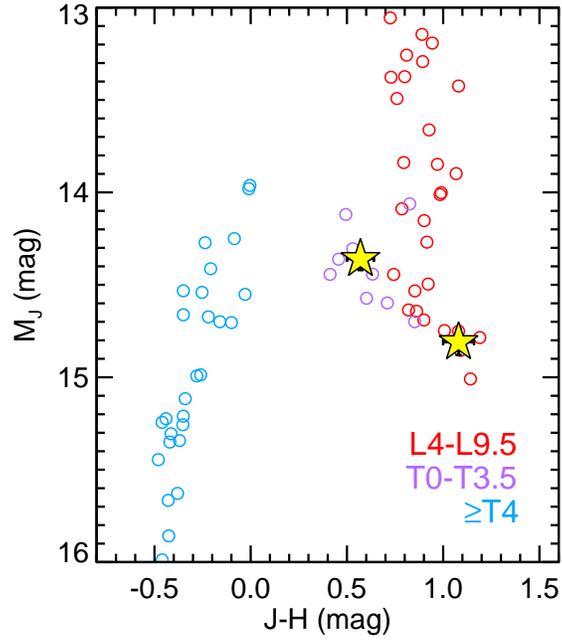}}

\caption{\normalsize Color--magnitude diagram showing the components
  of \obj{AB} (yellow stars) along with field L and T dwarfs with
  measured distances (open circles).  Both components have typical
  colors and magnitudes for their spectral types (L$6.5\pm1.5$ and
  T$1.5\pm1.0$).  Photometry is on the MKO system.  Field dwarf data
  were obtained from the Database of Ultracool Parallaxes
  (\url{http://www.as.utexas.edu/~tdupuy/plx/};
  \citealp{2012ApJS..201...19D}), and we only plot objects with
  uncertainties $<$10\% in parallax and $<$0.10\,mag in
  color. \label{fig:field-cmd}}

\end{figure}

Table~\ref{tbl:meas} provides a summary of all of the directly
measured properties of the \obj{AB} system.
Figure~\ref{fig:field-cmd} shows the components of \obj{AB} on a
color--magnitude diagram in comparison to other field L and T dwarfs
with measured distances.

%----------------------------------------------------------------------%

\section{Model-Derived Properties for \obj{AB} \label{sec:model}}

With a precisely determined total dynamical mass (6\%), component
masses (7\%), and component luminosities (15\%--20\%), we can derive
all other physical properties (\Teff, \logg, age, etc.) by invoking
evolutionary models.
Only one set of models currently incorporates cloud dispersal at the
L/T transition, which is particularly relevant for \obj{AB}.  SM08
``hybrid'' models assume the photosphere smoothly transitions from
cloudy
%($f_{\rm sed} = 2$) 
to cloudless as objects cool from effective temperatures of 1400\,K to
1200\,K.
Because \obj{A} is expected to be cloudy based on its late-L spectral
type, and \obj{B} likely still possesses some cloud opacity at the
photosphere, we also consider the SM08 fully cloudy ($f_{\rm sed} =
2$) and Lyon Dusty \citep{2000ApJ...542..464C} models.

To derive model properties from the individual masses and luminosities
only requires a straightforward bilinear interpolation of model
tracks.  But this could result in very different ages if models do not
accurately predict the mass--luminosity relation for our objects.
Because we are also interested in deriving properties under the
assumption of coevality, we also use our (more precise) total mass and
individual luminosities, ignoring our measured mass ratio, to derive
properties from evolutionary models in the same fashion as in our
previous work \citep{2008ApJ...689..436L, 2009ApJ...692..729D}. In
this coeval analysis, at each point on a log(age) grid we use the
luminosity of each component to calculate their model-predicted mass,
\Teff, surface gravity, radius, lithium abundance, and near-infrared
colors. This is done in a Monte Carlo fashion such that we use 10$^3$
values for a component's \Lbol, resulting in 10$^3$ mass estimates at
each age. We then step through each of these 10$^3$ \Lbol\ pairs,
considering the full range of ages for that pair, sum the component
masses as a function of age, and determine the age that matches the
measured total mass by interpolating the curve. This is also done in a
Monte Carlo fashion by repeating this step 10$^3$ times using randomly
drawn values for the measured \Mtot\ from our MCMC posterior. This
results in 10$^6$ model-derived values for every parameter and
accounts for the errors in both \Lbol\ and \Mtot\ while appropriately
tracking their covariances via the common uncertainty in the distance.

We report the median, 1$\sigma$, and 2$\sigma$ credible intervals of
the model-derived parameter distributions in the case where we used
the individual masses and in the case where we used the total mass
assuming coevality (Table~\ref{tbl:derived}).

\subsection{System Age \label{sec:age}}

One of the fundamental predictions of substellar evolutionary models
is how luminosity changes with age for a given mass (or changes with
mass at a given age). Thus, by measuring the component masses and
luminosities of \obj{AB} we can test whether models successfully give
the same age for the two components. (By a typical field age of
$\sim$1--10\,Gyr, even large differences in formation time of a few
Myr would result in binaries that are coeval to $\sim$0.001\,dex.) We
can also assume that the age is the same and use the individual
luminosities and total mass, ignoring our mass ratio, to derive a
single best matching model-derived age.

First, we test the widely used, fully cloudy models for coevality.
Given our individually measured masses and luminosities, Lyon Dusty
models give ages of $1.01^{+0.15}_{-0.17}$\,Gyr and
$0.66^{+0.10}_{-0.12}$\,Gyr for the primary and secondary of \obj{AB},
respectively.  Accounting for the covariance in distance and mass
ratio, the age difference is $\Delta\log{t} = 0.19\pm0.10$\,dex,
2.0$\sigma$ discrepant with being coeval.  The SM08 cloudy models give
similar ages to Lyon Dusty but somewhat more coeval with
$\Delta\log{t} = 0.16\pm0.10$\,dex (1.6$\sigma$ different from
coeval). In contrast to both of these cases, the SM08 hybrid models
give ages consistent with coevality at 0.9$\sigma$, $\Delta\log{t} =
0.09\pm0.12$\,dex.

The more realistic assumption of SM08 hybrid models that clouds
disappear as temperatures cool from 1400\,K to 1200\,K results in
higher luminosities at a given mass and age during the transition.
This higher luminosity is not simply due to less cloud opacity. The
difference in entropy between a cloudy 1400\,K brown dwarf and a
cloudless 1200\,K brown dwarf is greater than the entropy difference
of two brown dwarfs at those temperatures that are both cloudy
\citep{2008ApJ...689.1327S}.  Therefore, luminosity evolution should
appear to slow down as brown dwarfs cool through the L/T transition
because it takes longer to shed this excess entropy, causing a phase
of increased luminosity compared to either cloudy or cloudless models.
This means that the mass--luminosity relation at a given age becomes
shallower in the L/T transition, so that a given luminosity ratio
could correspond to a mass ratio further from unity, like the one we
measured directly ($0.78\pm0.07$, Section~\ref{sec:mass}).  Therefore,
it is not surprising that the SM08 hybrid models give ages in better
agreement with coevality for \obj{AB}.

\begin{figure}

\centerline{
\includegraphics[width=3.2in,angle=0]{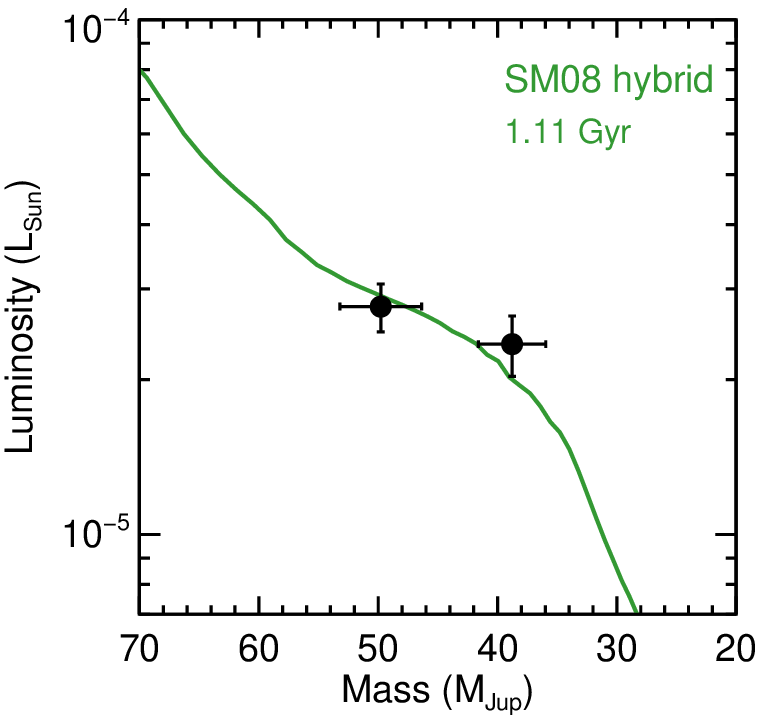} \hskip -0.4in
\includegraphics[width=3.2in,angle=0]{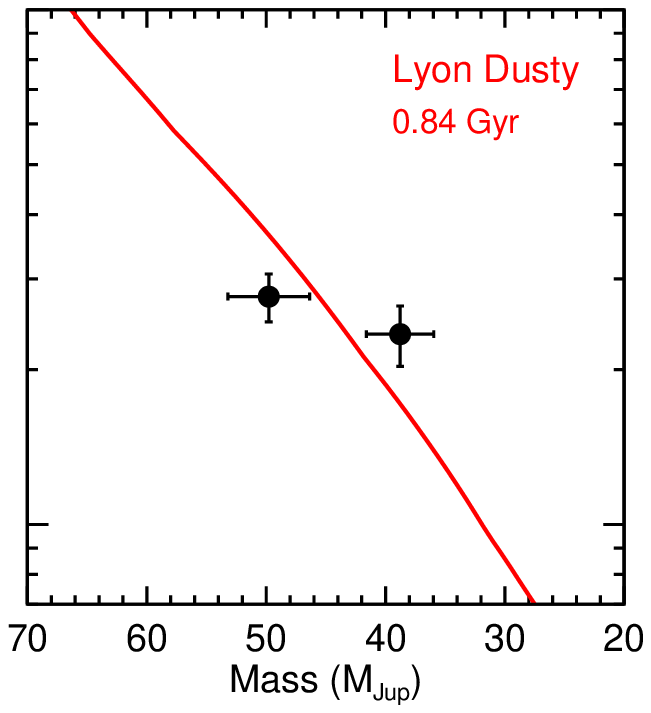}}

\caption{\normalsize Our directly measured individual masses and
  luminosities for the components of \obj{AB} compared to predictions
  from SM08 hybrid (left) and Lyon Dusty (right) evolutionary models.
  Model tracks are shown for the single coeval system age that best
  matches the total mass and individual luminosities. The unexpectedly
  shallow mass--luminosity relation implied by our data are better
  described by the SM08 hybrid models that show a slowing of
  luminosity evolution for objects in the L/T transition, while Lyon
  Dusty models are inconsistent with coevality at 2.0$\sigma$.  (Note
  that we do not plot a confidence range for models as that would
  effectively be double-plotting our errors, since the age of the
  plotted isochrone is derived from our observed total mass and
  component luminosities.)  \label{fig:m-l}}

\end{figure}

If we force coevality by ignoring our measured mass ratio, then we
find single best matching model-derived ages of
$1.11^{+0.17}_{-0.20}$\,Gyr (SM08 hybrid) and
$0.84^{+0.10}_{-0.15}$\,Gyr (Lyon Dusty).  Figure~\ref{fig:m-l} shows
the mass--luminosity relation predicted by models at these respective
coeval ages, illustrating the fundamental difference in the predicted
luminosity evolution between these two models. Over the mass range
40--50\,\Mjup, the Lyon Dusty isochrone has a power-law slope of
$\Delta\log{\Lbol}/\Delta\log{M} = 3.1$.  In contrast, for the SM08
hybrid models this slope is only 1.3.  Our directly measured masses
for \obj{AB} imply a power-law slope $\Delta\log{\Lbol}/\Delta\log{M}
= 0.6^{+0.6}_{-0.8}$ over the same $\approx$40--50\,\Mjup\ mass range.
Thus, we find a mass--luminosity relation in the L/T transition that
is in much better agreement with SM08 hybrid models than fully cloudy
models. In fact, our slope seems to be even shallower than the hybrid
models and is even nominally consistent with a inverted relation
($\Delta\log{\Lbol}/\Delta\log{M} < 0$) within the 1$\sigma$
uncertainty.

Finally, we note that another way of framing the coevality test is to
compare the model-derived mass ratios with our observed value of
$0.78\pm0.07$. When using just our total dynamical mass and individual
luminosities, both cloudy models give similar mass ratios of
$0.94^{+0.05}_{-0.06}$ (SM08) and $0.94\pm0.05$ (Lyon). These are much
closer to unity than we observe because the steeper mass--luminosity
relation predicted by cloudy models gives a very small difference in
mass for a correspondingly small difference in observed luminosity
($\Delta\log{\Lbol} = 0.07\pm0.07$\,dex). In comparison, SM08 hybrid
models predict a mass ratio of $0.87^{+0.11}_{-0.09}$ that is somewhat
larger than but consistent with our measured value at 0.9$\sigma$.

\subsection{Effective Temperature \& Surface Gravity \label{sec:teff}}

Combining evolutionary model radii with a measured luminosity and mass
readily produces estimates of effective temperature ($\Teff \propto
\Lbol^{-1/4} R^{-1/2}$) and surface gravity ($g \propto M R^{-2}$),
respectively. There are only small differences between the radii
predicted at a given age by the models considered here
($\lesssim$3\%), resulting in differences of $\lesssim$1\% in \Teff\
and $\lesssim$0.03\,dex in \logg. More important to the model-derived
radii is whether we force coevality, in which case the secondary is
predicted to be only slightly larger ($\leq$0.5\%) than the primary.
When the two components are allowed to have different ages but correct
masses, the model-derived age of the secondary is 0.1--0.3\,dex
younger (1$\sigma$ range), so its predicted radius is 3\%--6\%
larger. Therefore, we adopt the coeval model-derived temperatures and
surface gravities of \obj{AB}, using the SM08 hybrid models that are
most consistent with coevality.

The model-derived temperature of the L$6.5\pm1.5$ spectral type
primary is $1330\pm30$\,K, while the T$1.5\pm1.0$ secondary is
$1270^{+40}_{-30}$\,K. Their predicted surface gravities are $\logg =
5.10^{+0.05}_{-0.04}$\,dex and $5.04^{+0.05}_{-0.04}$\,dex,
respectively.  Interestingly, the mean evolutionary model-derived
temperature of the two components ($\approx$1300\,K in both coeval and
non-coeval cases) is in excellent agreement with the atmospheric model
fitting results of the integrated-light 1--14.5\,\micron\ spectrum of
\obj{AB} from \citet{2009ApJ...702..154S} who found $\Teff = 1300$\,K
(acceptable range of 1200--1400\,K) and $\logg = 5.5$\,dex
(5.0--5.5\,dex). A similar agreement between evolutionary and
atmospheric model temperatures has been seen for the only other L/T
transition brown dwarf with a dynamical mass determination
\citep[LHS~2397aB, $T_{\rm eff}^{\rm evol} = 1430\pm40$\,K and $T_{\rm
  eff}^{\rm atm} = 1400$\,K;][]{2009ApJ...699..168D}. Finally, we note
that the model-derived temperatures for \obj{AB} align very well with
the assumption made in SM08 hybrid models that the L/T transition
occurs over the temperature range 1200--1400\,K.

\subsection{Near-infrared Colors \label{sec:colors}}

\begin{figure}

\centerline{\includegraphics[width=2.5in,angle=0]{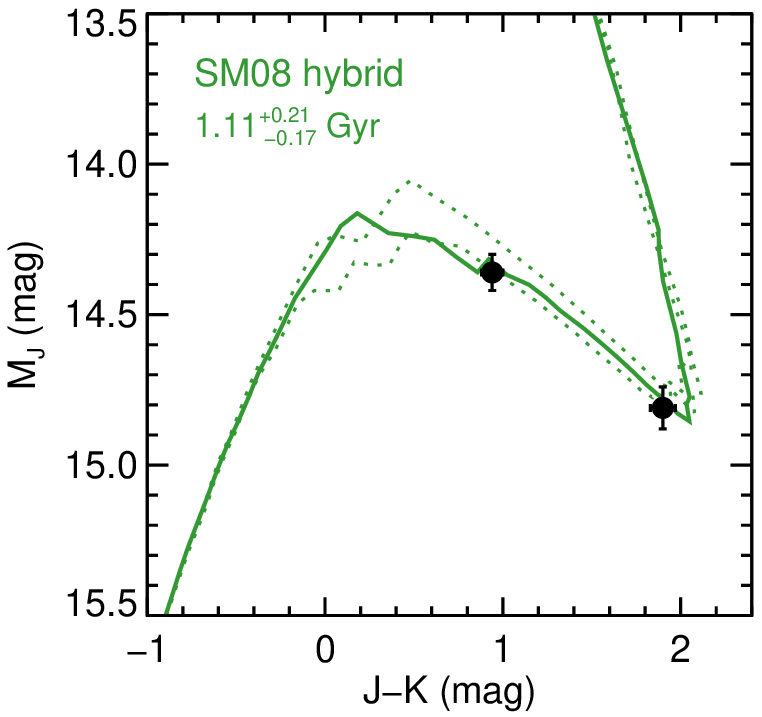}} \vskip 0.1in
\centerline{\includegraphics[width=2.5in,angle=0]{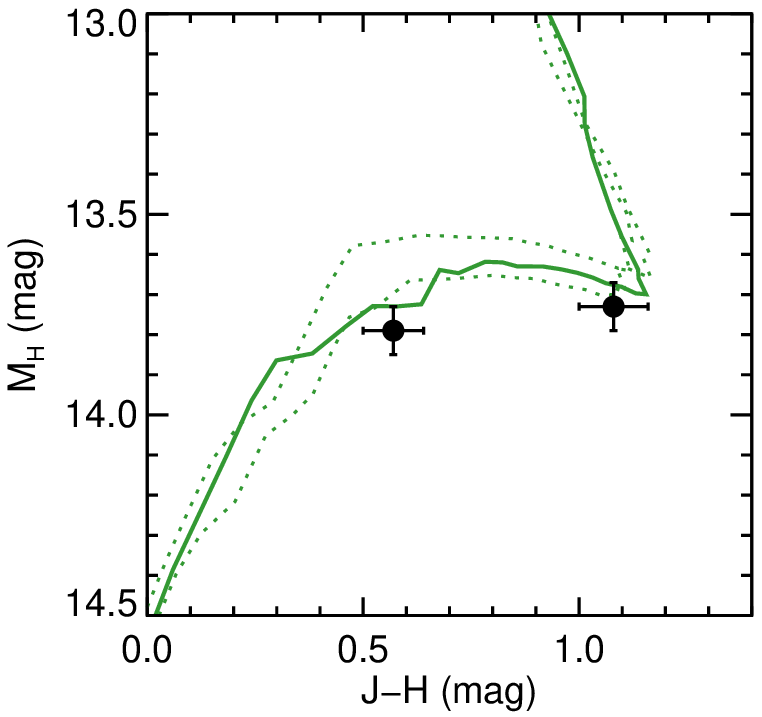}} \vskip 0.1in
\centerline{\includegraphics[width=2.5in,angle=0]{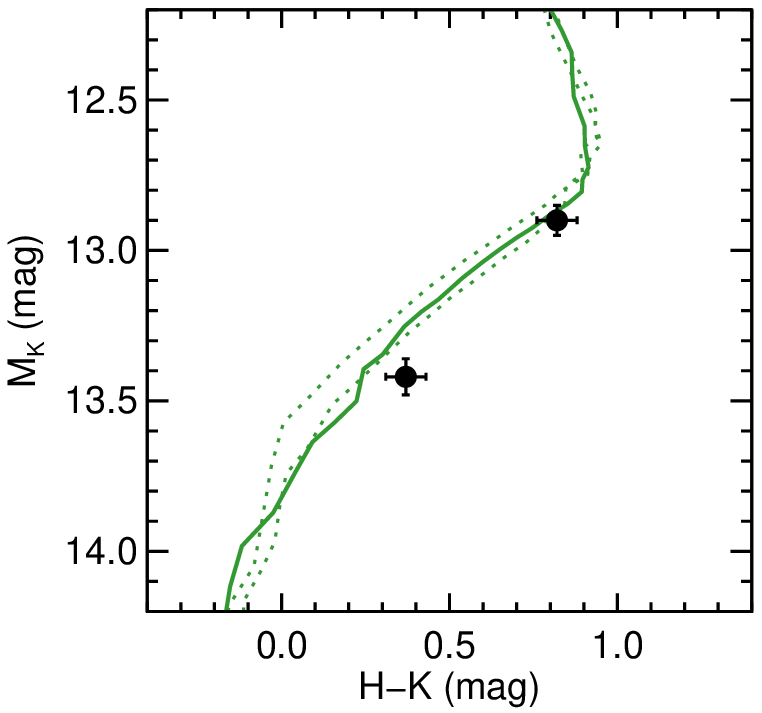}}

\caption{\normalsize Measured colors and absolute magnitudes of
  the components of \obj{AB} compared to predictions from SM08 hybrid
  evolutionary models.  Model tracks are shown for the coeval system
  age that best matches the total mass and individual luminosities
  (solid) and ages at $\pm$1$\sigma$ of this value (dotted). Unlike
  previous generations of evolutionary models, the predicted colors
  and magnitudes of SM08 hybrid match our observations remarkably
  well.  \label{fig:cmd}}

\end{figure}

We have independently measured the $JHK$ colors of the components of
\obj{AB} by combining our Keck flux ratios with the photometry from
\citet{2006AJ....131.2722C}. All colors agree within 1$\sigma$ of the
predictions of the SM08 hybrid models whether we enforce coevality or
not, although there is somewhat better agreement when deriving colors
directly from the individual masses and luminosities (non-coeval).
This agreement is remarkable as all other ultracool dwarfs with
dynamical mass determinations to date have typically shown
$\gtrsim$0.3\,mag disagreements with models
\citep[e.g.,][]{2008ApJ...689..436L, 2009ApJ...692..729D,
  2010ApJ...721.1725D, 2014ApJ...790..133D}. For example, the Lyon
Dusty models predict $\approx$3--4\,mag redder $J-K$ colors for the
components of \obj{AB}, which is not surprising given their assumption
of maximal dust clouds.  The reason that the SM08 hybrid models agree
with our observed $JHK$ colors is because these evolutionary models
also predict a $J$-band flux reversal for a system like \obj{AB}. The
model-derived flux ratios from the individual masses and luminosities
are $\Delta{J} = -0.50^{+0.15}_{-0.17}$\,mag and $\Delta{K} =
0.27^{+0.23}_{-0.25}$\,mag, which are quite similar to our measured
values ($\Delta{J} = -0.45\pm0.09$\,mag, $\Delta{K} =
0.52\pm0.05$\,mag).  Figure~\ref{fig:cmd} shows our observed colors
and magnitudes for \obj{AB} compared to SM08 hybrid evolutionary model
tracks.

\subsection{Lithium Depletion \label{sec:lith}}

According to \citet{1996ApJ...459L..91C}, most of the initial supply
of a $\geq$0.06\,\Msun\ brown dwarf's lithium is destroyed via fusion
by an age of $\leq$0.26\,Gyr, and 99\% is destroyed by
$\leq$1.00\,Gyr. The component masses of \obj{AB} are
$0.047\pm0.003$\,\Msun\ and $0.037^{+0.002}_{-0.003}$, so the Lyon
models predict that they should have retained almost all of their
lithium (${\rm Li}/{\rm Li}_0 \geq 0.55$ at 2$\sigma$ for the
primary).  However, even if both components of \obj{AB} are lithium
bearing, they may not possess a significant amount of atomic lithium
that would be readily detectable via the \ion{Li}{1} doublet at
6708~\AA.  At temperatures $\lesssim$1500\,K, most lithium in the
photosphere ($\approx$1\,bar) is predicted to be locked up in
molecular LiCl \citep{1999ApJ...519..793L}.  Thus, given the 2$\sigma$
upper limit on our model-derived temperature for \obj{A} ($\Teff <
1390$\,K) it is theoretically expected that both components of
\obj{AB} are chemically depleted in their atomic lithium. On the other
hand, a homogeneous analysis of L and T dwarf optical spectra by
\citet{2008ApJ...689.1295K} found that the occurrence of lithium
absorption is highest at L6--L7 spectral types, overlapping with the
L$6.5\pm1.5$ spectral type of \obj{A}.  Moreover,
\citet{2014ApJ...790...90F} detected lithium absorption in
WISE~J104915.57$-$531906.1B \citep[T0.5;][]{2013ApJ...772..129B}),
which is now the only T dwarf known to possess lithium.  Notably,
\citet{2010A&A...510A..99K} do not detect lithium in
$\epsilon$~Indi~Ba (T1), although this may be due to the fact that it
is massive enough to have depleted its lithium
\citep{2009AIPC.1094..509C}.  Thus, it is unclear whether \obj{AB}
would show evidence for atomic lithium in its integrated-light
spectrum.  High-resolution optical spectroscopy of \obj{AB} would
provide a unique, joint test of the theoretical lithium-fusing
mass-limit and atmospheric model predictions of the chemical depletion
of lithium.

%----------------------------------------------------------------------%

\section{Conclusions}

We have discovered that \obj{AB} (L6.5+T1.5) is a $J$-band flux
reversal binary.  We present precise individual dynamical masses by
combining resolved Keck AO orbital monitoring spanning 9.0\,yr with
integrated-light CFHT/WIRCam astrometric monitoring spanning 6.8\,yr,
the first such masses for any field L or T dwarfs. Despite spectral
types that are similar and luminosities that are indistinguishable
within the errors, we find a surprisingly low mass ratio of $q =
0.78\pm0.07$.  The only ultracool dwarf binary with a more precise
mass ratio is LHS~1070BC \citep[$0.92\pm0.01$][]{2012A&A...541A..29K},
also measured from astrometry, which highlights the greater potential
of astrometry for measuring precise individual masses as compared to
radial velocities.  For example, our mass ratio is based on a total of
only 2.4\,hr of integration time on a 4-m-class telescope, yet it is
more precise than the $q = 0.71^{+0.19}_{-0.13}$ measured for the
$\approx$5\,mag brighter binary Gl~569Bab from numerous resolved
spectroscopic observations from Keck \citep{2004astro.ph..7334O,
  2006ApJ...644.1183S, 2010ApJ...711.1087K}.  Combining our CFHT mass
ratio and Keck AO total mass gives component masses of
$49\pm3$\,\Mjup\ for \obj{A} (L$6.5\pm1.5$) and $39\pm3$\,\Mjup\ for
\obj{B} (T$1.5\pm1.0$).

This is the first $J$-band flux reversal binary or high-amplitude
variable with a dynamical mass measurement, providing a precise
benchmark for the cloud dispersal phase of substellar evolution. We
validate that the component fainter in $J$ band is in fact more
massive and that both components are unambiguously substellar
($<$75\,\Mjup). Perhaps the most striking result is the shallow
mass--luminosity relation in the L/T transition implied by our data
($\Delta\log{\Lbol}/\Delta\log{M} = 0.6^{+0.6}_{-0.8}$ over
$\approx$40--50\,\Mjup).  This disagrees with the mass--luminosity
relation predicted by fully cloudy models, providing the first direct
observational support that cloud dispersal plays an important role in
luminosity evolution.  We quantify this as a coevality test using our
measured individual masses and luminosities to derive an age from
evolutionary models for each component and test if the models
successfully give the same age for both components.  Lyon Dusty models
give ages that are different by $0.19\pm0.10$\,dex, a 2.0$\sigma$
discrepancy.  In comparison, hybrid models from
\citet{2008ApJ...689.1327S}, in which the dispersal of clouds causes a
slowing of luminosity evolution, gives component ages different by
$0.09\pm0.12$\,dex and thus consistent at (0.9$\sigma$).

In fact, these SM08 hybrid evolutionary models paint a remarkably
self-consistent picture for the properties of \obj{AB}. The models
assume that clouds disperse as temperatures cool from 1400\,K to
1200\,K. From our measured luminosities and SM08 model-derived radii
we find $\Teff = 1330\pm30$\,K for the L$6.5\pm1.5$ primary and
$1270^{+40}_{-30}$\,K for the T$1.5\pm1.0$ secondary. SM08 hybrid
models also accurately predict the $JHK$ colors of the components,
including the reversal in flux ratio observed between $J$ and $K$
bands. In addition, the \Teff\ of 1300\,K found for \obj{AB} by
\citet{2009ApJ...702..154S}, who used the same atmospheres in their
spectral synthesis modeling as are used by SM08 evolutionary models,
is in excellent agreement with our temperatures derived from
luminosities and model radii.  We note that without an independent
measurement of the age of \obj{AB}, we cannot rule out a constant
systematic offset in the SM08 hybrid model luminosities, as our
coevality test only constrains slope of the mass--luminosity relation.
For example, mid-L dwarfs appear to be 0.2--0.4\,dex more luminous
than predicted by models at a given mass and age
\citep{2009ApJ...692..729D, 2014ApJ...790..133D}.  If this holds true
for L/T transition objects, then the age we derive from SM08 models
would be underestimated by a factor of $\approx$2--3.

Overall, it seems that the distinguishing features \obj{AB}, like a
$J$-band flux reversal and high-amplitude variability, are normal for
a field L/T binary caught during the process of cloud dispersal.
\obj{AB}'s model-derived age of $1.11^{+0.17}_{-0.20}$\,Gyr is typical
of field brown dwarfs \citep[e.g.,][]{2007ApJ...666.1205Z}, and the
component surface gravities are correspondingly unexceptional, $\logg
= 5.0$--5.2\,dex.  The one unexpected physical property is the low
mass ratio.  To determine if this is a typical feature of L/T
transition binaries, especially for $J$-band flux reversal systems,
will require more individual mass measurements for late-L to early-T
type brown dwarfs.  Fortunately, such masses will likely be available
in the near future as our CFHT astrometric monitoring continues. Orbit
determinations typically require $\approx$30\% coverage of the orbital
period, and we have been obtaining CFHT data on our Keck dynamical
mass sample for $\approx$8\,yr.  Thus, L/T binaries with orbital
periods $\lesssim$20\,yr should soon have photocenter semimajor axis
measurements that will enable precise individual dynamical masses to
further map out the substellar mass--luminosity relation.

Our results lend further support to the growing evidence that clouds
have a significant impact on the luminosity evolution of substellar
objects. A shallow mass--luminosity relation in the L/T transition
suggests that even when the age and luminosity of an object are
constrained its mass may be difficult to estimate precisely.  This
adds another obstacle to estimating masses for directly imaged
extrasolar planets in this spectral type range \citep[e.g.,
HR~8799b;][]{2010ApJ...723..850B, 2011ApJ...733...65B}. The L/T
transition corresponds to the breakup of mostly silicate and iron
clouds.  At cooler temperatures, clouds composed of sulfides emerge
\citep[$\Teff \lesssim 900$\,K;][]{2012ApJ...756..172M} and water ice
clouds possibly at $\lesssim$350\,K \citep{2014ApJ...787...78M}. Even
though sulfide clouds are expected to be thinner, in principle they
could impact luminosity evolution in a comparable way as we have now
observed for silicate clouds, implying similar alterations to the
mass--luminosity relation for much colder brown dwarfs. Directly
measured individual masses for late-T and Y~dwarf binaries should be
able to test this idea.

%----------------------------------------------------------------------%

\acknowledgments

We thank the referee for a timely and very helpful review.
It is a pleasure to thank Antonin Bouchez, David LeMignant, Marcos van
Dam, Randy Campbell, Gary Punawai, Peter Wizinowich, and the Keck
Observatory staff for their efforts assisting on our first LGS AO
night that resulted in the discovery of \obj{AB}.
We also thank
Joel Aycock, Al Conrad, Greg Doppmann, Heather Hershley, Jim Lyke,
Jason McIlroy, Julie Riviera, Hien Tran, and Cynthia Wilburn for
assistance with subsequent Keck LGS AO observing.
We greatly appreciate the CFHT staff for their constant observing
support and dedication to delivering the highest quality data products.
This work was supported by a NASA Keck PI Data Award, administered by
the NASA Exoplanet Science Institute. 
M.C.L.\ acknowledges support from NSF grant AST09-09222.
Our research has employed the 2MASS data products; NASA's
Astrophysical Data System; the SIMBAD database operated at CDS,
Strasbourg, France; and the SpeX Prism Spectral Libraries, maintained
by Adam Burgasser at \url{http://www.browndwarfs.org/spexprism}.
Finally, the authors wish to recognize and acknowledge the very
significant cultural role and reverence that the summit of Mauna Kea has
always had within the indigenous Hawaiian community.  We are most
fortunate to have the opportunity to conduct observations from this
mountain.

{\it Facilities:} \facility{Keck:II (LGS AO, NGS AO, NIRC2)},
\facility{CFHT (WIRCam)} \facility{IRTF (SpeX)}

\clearpage

\clearpage
\begin{landscape}
\begin{deluxetable}{lclccccc}
\tablecaption{Relative astrometry and photometry for \obj{AB} from Keck/NIRC2 AO \label{tbl:keck}}
\tablewidth{0pt}
\tablehead{
\colhead{Date} &
\colhead{Airmass} &
\colhead{Filter} &
\colhead{FWHM} &
\colhead{Strehl ratio} &
\colhead{$\rho$} &
\colhead{PA} &
\colhead{$\Delta{m}$} \\
\colhead{(UT)} &
\colhead{} &
\colhead{} &
\colhead{(mas)} &
\colhead{} &
\colhead{(mas)} &
\colhead{(\degree)} &
\colhead{(mag)} }
\startdata
2005~May~1    & 1.342 &\phn\phn \Kp & $53\pm2$ & $0.27\pm0.02$ & $42.4\pm0.7$ &\phn$ 81.1\pm1.7$  & $ 0.51\pm0.03$     \\
2006~May~5    & 1.544 &\phn\phn $J$ & $57\pm7$ &$0.045\pm0.008$& $69.2\pm0.9$ &    $112.2\pm0.7$  & $-0.61\pm0.11$\phs \\
2006~May~5    & 1.572 &\phn\phn $H$ & $51\pm3$ & $0.11\pm0.02$ & $70.5\pm1.4$ &    $112.2\pm0.7$  & $ 0.00\pm0.13$     \\
2006~May~5    & 1.274 &\phn\phn \Ks & $54\pm2$ & $0.32\pm0.06$ & $70.3\pm0.7$ &    $112.2\pm0.7$  & $ 0.49\pm0.05$     \\
2006~Dec~19   & 1.099 &\phn\phn $K$ & $66\pm6$ & $0.16\pm0.05$ & $79.8\pm1.9$ &    $124.5\pm1.4$  & $ 0.49\pm0.15$     \\
2007~Mar~8    & 1.102 &\phn\phn \Kp & $59\pm5$ & $0.20\pm0.04$ & $80.3\pm0.5$ &    $125.4\pm0.6$  & $ 0.45\pm0.06$     \\
2007~Mar~25   & 1.102 &\phn\phn $J$ & $52\pm6$ &$0.049\pm0.016$& $80.9\pm0.5$ &    $128.1\pm1.0$  & $-0.36\pm0.05$\phs \\
2008~Apr~1    & 1.582 &\phn\phn $K$ & $69\pm2$ & $0.16\pm0.05$ & $72.2\pm1.7$ &    $142.3\pm1.2$  & $ 0.42\pm0.07$     \\
2008~Nov~3    & 1.373 &\phn\phn $H$ & $57\pm4$ & $0.08\pm0.02$ & $55.2\pm1.0$ &    $160.2\pm2.1$  & $-0.01\pm0.07$\phs \\
2008~Dec~22   & 1.132 &\phn\phn \Kp & \nodata  &    \nodata    & $50.3\pm0.7$ &    $162.4\pm0.9$  & $ 0.51\pm0.03$     \\
2010~Jan~10   & 1.188 &\phn\phn $H$ & \nodata  &    \nodata    & $34.4\pm0.5$ &    $234.9\pm1.1$  & $ 0.14\pm0.03$     \\
2010~Jan~10   & 1.169 &\phn\phn $K$ & \nodata  &    \nodata    & $33.1\pm0.6$ &    $235.1\pm1.2$  & $ 0.57\pm0.05$     \\
2010~May~22   & 1.316 &\phn\phn $K$ & \nodata  &    \nodata    & $39.9\pm0.7$ &    $261.0\pm1.1$  & $ 0.56\pm0.04$     \\
2011~Apr~21   & 1.212 &\phn\phn $K$ & \nodata  &    \nodata    & $58.9\pm0.3$ &   $297.91\pm0.29$ & $0.545\pm0.016$    \\
2014~May~10   & 1.326 &\phn\phn $K$ & \nodata  &    \nodata    & $54.9\pm0.9$ &\phn$ 97.2\pm1.0$  & $ 0.51\pm0.04$     \\

\enddata

\tablecomments{For the Keck imaging data, Strehl ratios and FWHM were
  computed using the publicly available routine \texttt{NIRC2STREHL}.
  Masking observations have no FWHM and Strehl listed.}

\end{deluxetable}
\clearpage
\end{landscape}
\clearpage
\begin{deluxetable}{lcccccc}
\tablewidth{0pt}
\tablecaption{Integrated-light astrometry for \obj{AB} from CFHT/WIRCam \label{tbl:cfht}}

\tablehead{
\colhead{Date}   &
\colhead{R.A.}    &
\colhead{Dec.}      &
\colhead{$\sigma_{\rm R.A.}\cos{\delta}$} &
\colhead{$\sigma_{\rm Dec.}$}  &
\colhead{Airmass}      &
\colhead{Seeing}     \\
\colhead{(UT)}   &
\colhead{(deg)}    &
\colhead{(deg)}      &
\colhead{(mas)} &
\colhead{(mas)} &
\colhead{} &
\colhead{} }
\startdata
2008~Feb~17 & 163.05659646 & +44.38217395 &  1.8 &  2.3 & 1.100 & 0$\farcs$55 \\
2008~Feb~23 & 163.05659550 & +44.38217470 &  1.6 &  2.8 & 1.099 & 0$\farcs$59 \\
2008~Apr~19 & 163.05658483 & +44.38216886 &  2.1 &  1.5 & 1.104 & 0$\farcs$60 \\
2008~Apr~28 & 163.05658379 & +44.38216735 &  2.4 &  2.0 & 1.100 & 0$\farcs$53 \\
2009~Apr~15 & 163.05659324 & +44.38213225 &  1.1 &  2.0 & 1.101 & 0$\farcs$59 \\
2009~Jun~6  & 163.05658999 & +44.38212290 &  2.6 &  2.8 & 1.160 & 0$\farcs$62 \\
2010~Mar~24 & 163.05660297 & +44.38209950 &  1.6 &  1.7 & 1.107 & 0$\farcs$90 \\
2010~Apr~21 & 163.05659864 & +44.38209535 &  1.4 &  1.6 & 1.100 & 0$\farcs$62 \\
2010~May~5  & 163.05659707 & +44.38209378 &  1.2 &  1.5 & 1.099 & 0$\farcs$59 \\
2011~Feb~12 & 163.05662003 & +44.38206551 &  1.5 &  2.3 & 1.118 & 0$\farcs$62 \\
2011~Mar~20 & 163.05661377 & +44.38206406 &  3.2 &  2.3 & 1.128 & 0$\farcs$67 \\
2011~Apr~15 & 163.05660884 & +44.38206101 &  2.3 &  3.0 & 1.122 & 0$\farcs$83 \\
2011~May~18 & 163.05660425 & +44.38205434 &  2.1 &  3.8 & 1.125 & 0$\farcs$66 \\
2011~Dec~9  & 163.05663748 & +44.38202911 &  2.0 &  1.9 & 1.099 & 0$\farcs$55 \\
2012~Jan~2  & 163.05663574 & +44.38202885 &  2.1 &  2.2 & 1.107 & 0$\farcs$59 \\
2012~Apr~4  & 163.05662005 & +44.38202585 &  1.6 &  2.1 & 1.114 & 0$\farcs$63 \\
2013~Apr~27 & 163.05663104 & +44.38198416 &  2.2 &  2.9 & 1.112 & 0$\farcs$58 \\
2013~Dec~25 & 163.05666361 & +44.38195306 &  2.7 &  2.1 & 1.107 & 0$\farcs$66 \\
2014~May~10 & 163.05664133 & +44.38194490 &  1.5 &  1.8 & 1.116 & 0$\farcs$54 \\
2014~Dec~2  & 163.05667169 & +44.38191431 &  2.5 &  1.8 & 1.104 & 0$\farcs$57 \\
2014~Dec~3  & 163.05667208 & +44.38191378 &  2.6 &  2.5 & 1.121 & 0$\farcs$55 \\
\enddata
\end{deluxetable}
\clearpage
\begin{deluxetable}{lcccc}
\setlength{\tabcolsep}{0.050in}
\tabletypesize{\small}
\tablewidth{0pt}
\tablecaption{Derived orbital and parallax parameters for \obj{AB} \label{tbl:orbit}}

\tablehead{
\colhead{Parameter}   &
\colhead{Best fit}    &
\colhead{Median}      &
\colhead{68.3\% c.i.} &
\colhead{95.4\% c.i.} }
\startdata
\multicolumn{5}{c}{Visual binary orbital parameters} \\
\cline{1-5}
Orbital period $P$ (yr)                                        & 8.614         & 8.608         &        8.583, 8.632        &        8.560, 8.658        \\
Semimajor axis $a$ (mas)                                       & 70.59         & 70.67         &        70.43, 70.91        &        70.20, 71.16        \\
Eccentricity $e$                                               & 0.1387        & 0.1399        &       0.1376, 0.1422       &       0.1354, 0.1445       \\
Inclination $i$ (\degree)                                      & 62.0          & 62.1          &         61.7, 62.4         &         61.4, 62.7         \\
PA of the ascending node $\Omega$ (\degree)                    & 126.7         & 126.8         &        126.5, 127.2        &        126.2, 127.5        \\
Argument of periastron $\omega$ (\degree)                      & 186.5         & 187.3         &        185.6, 188.9        &        184.0, 190.5        \\
Mean longitude at 2455197.5~JD $\lambda_{\rm ref}$ (\degree)   & 113.4         & 113.4         &        112.9, 113.8        &        112.5, 114.2        \\
\cline{1-5}
\multicolumn{5}{c}{} \\                                                                                                                                        
\multicolumn{5}{c}{Additional integrated-light astrometric parameters} \\
\cline{1-5}
${\rm R.A.} -163.0566182$ (mas)                                                    & 0.0           & $-$0.3\phs    &       $-$1.6, 0.9\phs      &       $-$2.8, 2.1\phs      \\
${\rm Dec.} -+44.3821006$ (mas)                                                    & 0.0           & 0.1           &       $-$0.5, 0.7\phs      &       $-$1.1, 1.3\phs      \\
Relative proper motion in R.A.\ $\mu_{\rm R.A., rel}$ (\masyr) & 24.51         & 24.56         &        24.36, 24.77        &        24.16, 24.97        \\
Relative proper motion in Dec.\ $\mu_{\rm Dec., rel}$ (\masyr) & $-$133.96     & $-$133.91     &    $-$134.14, $-$133.69    &    $-$134.37, $-$133.45    \\
Relative parallax $\pi_{\rm rel}$ (mas)                        & 36.87         & 36.67         &        36.06, 37.29        &        35.42, 37.90        \\
Photocenter semimajor axis $\alpha$ (mas)                      & $-$11.7       & $-$11.6       &      $-$12.2, $-$11.0      &      $-$12.8, $-$10.5      \\
\enddata

\tablecomments{For each parameter we report the value corresponding to
  the best fit (i.e., the lowest $\chi^2$ in the MCMC chain,
  $\chi^2_{\rm min} = 50.7$, 59 degrees of freedom) along with the
  median of the posterior distribution and the shortest intervals
  containing 68.3\% and 95.4\% of the chain steps (i.e., 1$\sigma$ and
  2$\sigma$ credible intervals). The time of periastron passage
  corresponding to these $\lambda_{\rm ref}$ and $\omega$ posteriors
  is $T_0 = 55842\pm13$~MJD (2011~Oct~7~UT).  For clarity, the R.A.\
  and Dec.\ zero points are reported relative to their best-fit
  values.  R.A.\ and Dec.\ zero points are reported at equinox J2000.0
  and epoch 2010.0.  Without resolved radial velocities there is a
  180\degree\ ambiguity in $\Omega$, $\omega$, and $\lambda_{\rm
    ref}$.}

\end{deluxetable}
\clearpage
\begin{deluxetable}{lccc}
\tablewidth{0pt}
\tablecaption{Measured Properties of \obj{AB} \label{tbl:meas}}
\tablehead{
\colhead{Property}   &
\colhead{\obj{A}} &
\colhead{\obj{B}} &
\colhead{Ref.}       }
\startdata
$d$ (pc)                   &            \multicolumn{2}{c}{$  26.1\pm0.5   $\phn        } &   1   \\
Semimajor axis (AU)        &            \multicolumn{2}{c}{$ 1.84^{+0.04}_{-0.03} $     } &   1   \\
\Mtot\ (\Mjup)             &            \multicolumn{2}{c}{$    88\pm5     $\phn       } &   1   \\
$q \equiv M_B/M_A$         &            \multicolumn{2}{c}{$  0.78\pm0.07   $           } &   1   \\
Mass (\Mjup)               &\phs    $        49\pm3    $&\phs    $        39\pm3     $    &  1,2  \\
Spectral type              &\phs    ${\rm L6.5}\pm1.5  $&\phs    ${\rm T1.5}\pm1.0   $    &   1   \\
$J$ (mag)                  &\phs    $    16.89\pm0.06  $&\phs    $     16.44\pm0.05  $    &  1,2  \\
$H$ (mag)                  &\phs    $    15.81\pm0.05  $&\phs    $     15.87\pm0.05  $    &  1,2  \\
$K$ (mag)                  &\phs    $    14.99\pm0.04  $&\phs    $     15.50\pm0.04  $    &  1,2  \\
$J-H$ (mag)                &\phs\phn$     1.08\pm0.08  $&\phs\phn$      0.57\pm0.07  $    &  1,2  \\
$H-K$ (mag)                &\phs\phn$     0.82\pm0.06  $&\phs\phn$      0.37\pm0.06  $    &  1,2  \\
$J-K$ (mag)                &\phs\phn$     1.90\pm0.07  $&\phs\phn$      0.94\pm0.06  $    &  1,2  \\
$M_J$ (mag)                &\phs    $    14.81\pm0.07  $&\phs    $     14.36\pm0.06  $    &  1,2  \\
$M_H$ (mag)                &\phs    $    13.73\pm0.06  $&\phs    $     13.79\pm0.06  $    &  1,2  \\
$M_K$ (mag)                &\phs    $    12.90\pm0.05  $&\phs    $     13.42\pm0.06  $    &  1,2  \\
BC$_K$ (mag)               &\phs\phn$     3.25\pm0.10  $&\phs\phn$      2.91\pm0.13  $    &  1,3  \\
$\log$(\Lbol/\Lsun)        &\phn    $    -4.56\pm0.05  $&\phn    $     -4.63\pm0.06  $    &   1   \\
$\Delta\log$(\Lbol)        &            \multicolumn{2}{c}{$  0.07\pm0.07  $            } &   1   \\
Parallax (mas)             &            \multicolumn{2}{c}{$  38.4\pm0.7   $\phn        } &   1   \\
$\mu_{\rm R.A.}$ (\masyr)  &            \multicolumn{2}{c}{$   +19\pm3     $\phn\phs    } &   1   \\
$\mu_{\rm Dec.}$ (\masyr)  &            \multicolumn{2}{c}{$  -140\pm3     $\phn\phn\phs} &   1   \\
\enddata

\tablecomments{All near-infrared photometry is on the MKO system.
  Parallax and proper motion have the following additive offsets
  applied to correct for the mean motion of our astrometric reference
  grid: $\Delta{\pi} = 1.7\pm0.3$\,mas, $\Delta{\mu_{\rm R.A.}} =
  -6\pm3$\,\masyr, $\Delta{\mu_{\rm Dec.}} = -7\pm3$\,\masyr.}

\tablerefs{(1)~This work; (2)~\citet{2006AJ....131.2722C}; (3)~\citet{2010ApJ...722..311L}.}

\end{deluxetable}
\clearpage
\begin{landscape}
\clearpage
\begin{deluxetable}{lcccccccccccc}
\setlength{\tabcolsep}{0.050in}
\tabletypesize{\small}
\tablewidth{0pt}
\tablehead{
\colhead{} & 
\colhead{} & \multicolumn{3}{c}{\citet{2008ApJ...689.1327S} hybrid} &
\colhead{} & \multicolumn{3}{c}{SM08 cloudy ($f_{\rm sed} = 2$)} &
\colhead{} & \multicolumn{3}{c}{Lyon Dusty \citep{2000ApJ...542..464C}} \\
\cline{3-5}
\cline{7-9}
\cline{11-13}
\colhead{Property}    &
\colhead{} & 
\colhead{Median}      &
\colhead{68.3\% c.i.} &
\colhead{95.4\% c.i.} &
\colhead{}            &
\colhead{Median}      &
\colhead{68.3\% c.i.} &
\colhead{95.4\% c.i.} &
\colhead{}            &
\colhead{Median}      &
\colhead{68.3\% c.i.} &
\colhead{95.4\% c.i.} }
\tablecaption{Evolutionary model-derived properties for \obj{AB} \label{tbl:derived}}
\startdata
\multicolumn{13}{c}{Using individual masses and luminosities} \\
\cline{1-13}
$t_{\rm A}$ (Gyr)              & & 1.22       &      0.99, 1.43      &      0.82, 1.69      & & 0.95       &      0.79, 1.07      &      0.69, 1.30      & & 1.01       &      0.84, 1.16      &      0.72, 1.36      \\
$t_{\rm B}$ (Gyr)              & & 0.99       &      0.79, 1.17      &      0.63, 1.39      & & 0.66       &      0.55, 0.77      &      0.45, 0.90      & & 0.66       &      0.54, 0.76      &      0.45, 0.90      \\
${\log(t_{\rm A}/{\rm yr})}$   & & 9.09       &      9.01, 9.17      &      8.93, 9.24      & & 8.98       &      8.91, 9.04      &      8.86, 9.12      & & 9.01       &      8.94, 9.08      &      8.87, 9.14      \\
${\log(t_{\rm B}/{\rm yr})}$   & & 9.00       &      8.92, 9.09      &      8.82, 9.16      & & 8.82       &      8.75, 8.89      &      8.67, 8.96      & & 8.82       &      8.74, 8.89      &      8.67, 8.97      \\
$\Delta\log{t}$ (dex)          & & 0.09       &   $-$0.03, 0.21\phs  &   $-$0.15, 0.33\phs  & & 0.16       &      0.06, 0.26      &   $-$0.03, 0.36\phs  & & 0.19       &      0.09, 0.29      &   $-$0.01, 0.39\phs  \\
$T_{\rm eff, A}$ (K)           & & 1340       &      1310, 1370      &      1280, 1400      & & 1320       &      1280, 1360      &      1250, 1400      & & 1360       &      1320, 1390      &      1280, 1430      \\
$T_{\rm eff, B}$ (K)           & & 1270       &      1230, 1300      &      1200, 1330      & & 1240       &      1190, 1270      &      1160, 1320      & & 1260       &      1220, 1300      &      1180, 1340      \\
$\Delta\Teff$ (K)              & & 70         &    \phn30, 110       & \phn$-$10, 150\phs   & & 90         &    \phn40, 140       & \phn$-$10, 180\phs   & & 100        &    \phn50, 150       & \phn\phn0, 190       \\
$\log(g_{\rm A})$ (cgs)        & & 5.14       &      5.10, 5.18      &      5.05, 5.22      & & 5.12       &      5.07, 5.16      &      5.03, 5.21      & & 5.15       &      5.10, 5.20      &      5.05, 5.24      \\
$\log(g_{\rm B})$ (cgs)        & & 5.00       &      4.96, 5.05      &      4.92, 5.09      & & 4.96       &      4.91, 5.01      &      4.86, 5.05      & & 4.99       &      4.94, 5.03      &      4.89, 5.08      \\
$R_{\rm A}$ (\Rjup)            & & 0.947      &     0.929, 0.965     &     0.912, 0.983     & & 0.970      &     0.953, 0.987     &     0.934, 1.004     & & 0.939      &     0.923, 0.957     &     0.901, 0.973     \\
$R_{\rm B}$ (\Rjup)            & & 0.972      &     0.950, 0.991     &     0.934, 1.017     & & 1.023      &     1.003, 1.041     &     0.985, 1.063     & & 0.991      &     0.972, 1.010     &     0.955, 1.030     \\
(Li/Li$_0$)$_{\rm A}$          & & \nodata    &       \nodata        &       \nodata        & & \nodata    &       \nodata        &       \nodata        & & 0.947      &     0.930, 0.986     &     0.545, 1.000     \\
(Li/Li$_0$)$_{\rm B}$          & & \nodata    &       \nodata        &       \nodata        & & \nodata    &       \nodata        &       \nodata        & & 1.0        &       1.0, 1.0       &       1.0, 1.0       \\
$(Y-J)_{\rm A}$ (mag)          & & 1.205      &     1.200, 1.212     &     1.187, 1.214     & & 1.201      &     1.194, 1.212     &     1.176, 1.215     & & \nodata    &       \nodata        &       \nodata        \\
$(Y-J)_{\rm B}$ (mag)          & & 1.182      &     1.165, 1.206     &     1.134, 1.215     & & 1.16       &      1.13, 1.18      &      1.12, 1.21      & & \nodata    &       \nodata        &       \nodata        \\
$(J-H)_{\rm A}$ (mag)          & & 1.04       &      0.94, 1.14      &      0.86, 1.21      & & 0.98       &      0.90, 1.13      &      0.71, 1.16      & & 2.51       &      2.41, 2.61      &      2.31, 2.71      \\
$(J-H)_{\rm B}$ (mag)          & & 0.73       &      0.57, 0.91      &      0.37, 1.02      & & 0.55       &      0.32, 0.69      &      0.26, 0.92      & & 2.78       &      2.68, 2.90      &      2.56, 3.00      \\
$(H-K)_{\rm A}$ (mag)          & & 0.71       &      0.56, 0.87      &      0.44, 0.99      & & 0.62       &      0.48, 0.87      &      0.26, 0.94      & & 2.05       &      1.96, 2.15      &      1.85, 2.25      \\
$(H-K)_{\rm B}$ (mag)          & & 0.29       &      0.09, 0.50      &   $-$0.12, 0.65\phs  & & 0.09       &   $-$0.17, 0.22\phs  &   $-$0.18, 0.53\phs  & & 2.33       &      2.26, 2.42      &      2.14, 2.48      \\
$(J-K)_{\rm A}$ (mag)          & & 1.8        &       1.5, 2.0       &       1.3, 2.2       & & 1.6        &       1.4, 2.0       &       1.0, 2.1       & & 4.56       &      4.36, 4.75      &      4.16, 4.96      \\
$(J-K)_{\rm B}$ (mag)          & & 1.0        &       0.7, 1.4       &       0.2, 1.7       & & 0.6        &       0.1, 0.9       &       0.1, 1.5       & & 5.11       &      4.93, 5.32      &      4.69, 5.48      \\
$(K-\Lp)_{\rm A}$ (mag)        & & 1.25       &      1.21, 1.29      &      1.18, 1.34      & & 1.28       &      1.22, 1.32      &      1.18, 1.39      & & 1.92       &      1.82, 2.03      &      1.72, 2.13      \\
$(K-\Lp)_{\rm B}$ (mag)        & & 1.36       &      1.29, 1.42      &      1.25, 1.50      & & 1.42       &      1.35, 1.50      &      1.28, 1.57      & & 2.23       &      2.12, 2.35      &      1.98, 2.44      \\
\cline{1-13}
\multicolumn{13}{c}{} \\                                                                                                                                        
\multicolumn{13}{c}{} \\                                                                                                                                        
\multicolumn{13}{c}{Using total mass, individual luminosities, and assuming coevality} \\
\cline{1-13}
Age ($t$, Gyr)                 & & 1.11       &      0.91, 1.28      &      0.76, 1.49      & & 0.81       &      0.69, 0.91      &      0.60, 1.04      & & 0.84       &      0.69, 0.94      &      0.61, 1.09      \\
${\log(t/{\rm yr})}$           & & 9.04       &      8.97, 9.12      &      8.89, 9.18      & & 8.91       &      8.85, 8.97      &      8.79, 9.02      & & 8.92       &      8.86, 8.99      &      8.79, 9.05      \\
$M_{\rm A}$ (\Mjup)            & & 47         &        43, 51        &        40, 55        & & 45.8       &      42.8, 48.6      &      40.3, 51.7      & & 45.5       &      42.6, 48.2      &      40.0, 51.5      \\
$M_{\rm B}$ (\Mjup)            & & 41         &        38, 44        &        35, 48        & & 42.6       &      39.7, 45.2      &      37.5, 48.4      & & 43.1       &      40.4, 45.9      &      37.6, 48.7      \\
$q \equiv M_{\rm B}/M_{\rm A}$  & & 0.87       &      0.78, 0.98      &      0.67, 1.09      & & 0.93       &      0.87, 0.98      &      0.83, 1.06      & & 0.94       &      0.89, 0.99      &      0.85, 1.05      \\
$T_{\rm eff, A}$ (K)           & & 1330       &      1300, 1360      &      1270, 1400      & & 1310       &      1270, 1340      &      1240, 1380      & & 1340       &      1300, 1370      &      1270, 1410      \\
$T_{\rm eff, B}$ (K)           & & 1270       &      1240, 1310      &      1200, 1350      & & 1250       &      1210, 1290      &      1170, 1340      & & 1280       &      1230, 1320      &      1200, 1370      \\
$\Delta\Teff$ (K)              & & 60         & \phn\phn0, 100       & \phn$-$50, 160\phs   & & 60         &    \phn20, 120       & \phn$-$50, 160\phs   & & 60         &    \phn10, 120       & \phn$-$50, 170\phs   \\
$\log(g_{\rm A})$ (cgs)        & & 5.10       &      5.06, 5.15      &      5.01, 5.19      & & 5.06       &      5.02, 5.10      &      4.98, 5.14      & & 5.09       &      5.04, 5.13      &      5.00, 5.18      \\
$\log(g_{\rm B})$ (cgs)        & & 5.04       &      5.00, 5.09      &      4.95, 5.13      & & 5.03       &      4.98, 5.06      &      4.95, 5.11      & & 5.06       &      5.02, 5.10      &      4.97, 5.14      \\
$R_{\rm A}$ (\Rjup)            & & 0.958      &     0.940, 0.974     &     0.924, 0.993     & & 0.991      &     0.975, 1.007     &     0.959, 1.022     & & 0.958      &     0.939, 0.969     &     0.930, 0.993     \\
$R_{\rm B}$ (\Rjup)            & & 0.960      &     0.942, 0.977     &     0.927, 0.997     & & 0.998      &     0.983, 1.013     &     0.967, 1.027     & & 0.963      &     0.944, 0.975     &     0.934, 0.997     \\
(Li/Li$_0$)$_{\rm A}$          & & \nodata    &       \nodata        &       \nodata        & & \nodata    &       \nodata        &       \nodata        & & 0.977      &     0.968, 1.000     &     0.943, 1.000     \\
(Li/Li$_0$)$_{\rm B}$          & & \nodata    &       \nodata        &       \nodata        & & \nodata    &       \nodata        &       \nodata        & & 0.992      &     0.984, 1.000     &     0.960, 1.000     \\
$(Y-J)_{\rm A}$ (mag)          & & 1.205      &     1.200, 1.213     &     1.185, 1.215     & & 1.200      &     1.191, 1.214     &     1.171, 1.216     & & \nodata    &       \nodata        &       \nodata        \\
$(Y-J)_{\rm B}$ (mag)          & & 1.184      &     1.167, 1.206     &     1.138, 1.214     & & 1.17       &      1.14, 1.19      &      1.12, 1.21      & & \nodata    &       \nodata        &       \nodata        \\
$(J-H)_{\rm A}$ (mag)          & & 1.01       &      0.90, 1.13      &      0.80, 1.20      & & 0.92       &      0.81, 1.08      &      0.65, 1.14      & & 2.57       &      2.47, 2.67      &      2.37, 2.76      \\
$(J-H)_{\rm B}$ (mag)          & & 0.77       &      0.60, 0.95      &      0.38, 1.07      & & 0.64       &      0.40, 0.81      &      0.29, 1.01      & & 2.73       &      2.62, 2.85      &      2.48, 2.96      \\
$(H-K)_{\rm A}$ (mag)          & & 0.67       &      0.48, 0.83      &      0.38, 0.98      & & 0.54       &      0.36, 0.73      &      0.18, 0.89      & & 2.10       &      2.01, 2.20      &      1.91, 2.29      \\
$(H-K)_{\rm B}$ (mag)          & & 0.3        &       0.1, 0.6       &    $-$0.1, 0.7\phs   & & 0.18       &   $-$0.10, 0.37\phs  &   $-$0.17, 0.65\phs  & & 2.26       &      2.16, 2.36      &      2.03, 2.44      \\
$(J-K)_{\rm A}$ (mag)          & & 1.7        &       1.4, 2.0       &       1.2, 2.2       & & 1.5        &       1.2, 1.8       &       0.8, 2.0       & & 4.67       &      4.48, 4.87      &      4.28, 5.03      \\
$(J-K)_{\rm B}$ (mag)          & & 1.1        &       0.7, 1.5       &       0.3, 1.8       & & 0.8        &       0.3, 1.2       &       0.1, 1.7       & & 5.0        &       4.8, 5.2       &       4.5, 5.4       \\
$(K-\Lp)_{\rm A}$ (mag)        & & 1.26       &      1.22, 1.30      &      1.18, 1.35      & & 1.29       &      1.24, 1.34      &      1.20, 1.40      & & 1.99       &      1.89, 2.09      &      1.78, 2.18      \\
$(K-\Lp)_{\rm B}$ (mag)        & & 1.35       &      1.28, 1.41      &      1.23, 1.50      & & 1.40       &      1.33, 1.49      &      1.25, 1.55      & & 2.16       &      2.04, 2.27      &      1.91, 2.37      \\
\enddata
\end{deluxetable}
\clearpage
\end{landscape}
\clearpage
\end{document}